\documentclass[12pt]{article}

\usepackage[utf8]{inputenc}
\usepackage{charter}
\usepackage{color}
\usepackage{fullpage}
\usepackage{hyperref}
\usepackage{textcomp}
\usepackage{amsmath}
\usepackage{hyperref}
\usepackage{graphicx}
\usepackage{upgreek}
\usepackage{wrapfig}
\usepackage{wrapfig,lipsum,booktabs}
\usepackage{lipsum}
\usepackage[sort&compress,numbers]{natbib}
\usepackage{hyperref}
\usepackage{soul,xcolor}
\usepackage[normalem]{ulem}

\usepackage[total={6.5in,9in},top=1in,headsep=0.1in,headheight=1in]{geometry}

\newcommand\snowmass{
\begin{center}
  \rule[-0.2in]{\hsize}{0.01in}\\
  \rule{\hsize}{0.01in}\\
  \vskip 0.1in
  Submitted to the Proceedings of the US Community Study\\ 
  on the Future of Particle Physics (Snowmass 2021)\\
  \rule{\hsize}{0.01in}\\
  \rule[+0.2in]{\hsize}{0.01in}\\[-2em]
\end{center}
}

\usepackage[firstpage=true]{background}
\backgroundsetup{contents={\parbox{6.5in}{\snowmass}}, scale=1,placement=top,opacity=1,color=black,position={3.25in,1.2in}}

\usepackage{fancyhdr}
\fancypagestyle{plain}{%
  \fancyhf{}%
  \fancyhead[C]{}
  \fancyfoot[C]{\thepage}
}

\fancypagestyle{empty}{%
  \fancyhf{}%
  \fancyhead[C]{{Snowmass2021 Cosmic Frontier: The landscape of cosmic-ray and high-energy photon probes of particle dark matter}}
  \fancyfoot[C]{\thepage}
}
\title{Snowmass2021 Cosmic Frontier: The landscape of cosmic-ray and high-energy photon probes of particle dark matter}
\date{}

\usepackage{authblk}

\newcommand{\til}{{\raise.17ex\hbox{$\scriptstyle\sim$}}}

\author[1]{Tsuguo Aramaki}
\affil[1]{Department of Physics, Northeastern University, Boston, MA 02115, USA}

\author[2,3]{Mirko Boezio}
\affil[2]{INFN, Sezione di Trieste, I-34149 Trieste, Italy}
\affil[3]{IFPU, I-34014 Trieste, Italy}

\author[4]{James Buckley}
\affil[4]{Department of Physics, Washington University, St. Louis, MO 63130, USA}

\author[5]{Esra Bulbul}
\affil[5]{Max Planck Institute for Extraterrestrial Physics, 85748 Garching, Germany}

\author[6]{Philip von Doetinchem}
\affil[6]{Department of Physics and Astronomy, University of Hawaii at Manoa, Honolulu, HI 96822, USA}

\author[7,8]{Fiorenza Donato}
\affil[7]{Dipartimento di Fisica, Università di Torino, 10125 Torino, Italy}
\affil[8]{Istituto Nazionale di Fisica Nucleare, Sezione di Torino, 10125 Torino, Italy}

\author[9]{J. Patrick Harding}
\affil[9]{Physics Division, Los Alamos National Laboratory, Los Alamos, NM, USA}

\author[10]{Chris Karwin}
\affil[10]{Department of Physics and Astronomy, Clemson University, Clemson, SC 29634, USA}

\author[6]{Jason Kumar}

\author[11,12]{Rebecca K. Leane}
\affil[11]{SLAC National Accelerator Laboratory, Menlo Park, CA 94025-7015, USA}
\affil[12]{Kavli Institute for Particle Astrophysics and Cosmology, Stanford University, Stanford, CA 94305-4085 USA}

\author[13]{Shigeki Matsumoto}
\affil[13]{Kavli IPMU (WPI), UTIAS,  University of Tokyo, Kashiwa 277-8583, Japan}

\author[14]{Julie McEnry}
\affil[14]{Astrophysics Science Division, NASA Goddard Space Flight Center, Greenbelt, MD 20771, USA}

\author[13]{Tom Melia}

\author[15]{Kerstin Perez}
\affil[15]{Department of Physics, Massachusetts Institute of Technology, Cambridge, MA 02139, USA}

\author[16]{Stefano Profumo}
\affil[16]{Department of Physics and Santa Cruz Institute for Particle Physics, University of California, Santa Cruz, CA 95064, USA}

\author[17]{Daniel Salazar-Gallegos}
\affil[17]{Department of Physics and Astronomy, Michigan State University, East Lansing, MI 48824, USA}

\author[5]{Andrew W. Strong}

\author[15]{Brandon Roach}

\author[18,19]{Miguel A. S\'anchez-Conde}
\affil[18]{Instituto de F\'isica Te\'orica UAM-CSIC, Universidad Aut\'onoma de Madrid, C/ Nicol\'as Cabrera, 13-15, 28049 Madrid, Spain}
\affil[19]{Departamento de F\'isica Te\'orica, M-15, Universidad Aut\'onoma de Madrid, E-28049 Madrid, Spain}

\author[11,12]{Tom Shutt}

\author[20]{Atsushi Takada}
\affil[20]{Graduate School of Science, Kyoto University, Kitashirakawa Oiwakecho, Sakyo, Kyoto, Kyoto, 606-8502, Japan}

\author[20]{Toru Tanimori}

\author[21]{John Tomsick}
\affil[21]{Space Sciences Laboratory, University of California, Berkeley, CA 94720, USA}

\author[13]{Yu Watanabe}

\author[16]{David A. Williams}

\begin{document}
\clubpenalty = 10000 \widowpenalty = 10000 \displaywidowpenalty = 10000

\maketitle

\newpage

\section*{Executive Summary}

This white paper discusses the current landscape and prospects for experiments sensitive to particle dark matter processes producing photons and cosmic rays. 
Much of the $\upgamma$-ray sky remains unexplored on a level of sensitivity that would enable the discovery of a dark matter signal.
Currently operating GeV--TeV observatories, such as Fermi-LAT, atmospheric Cherenkov telescopes, and water Cherenkov detector arrays continue to target several promising dark matter-rich environments within and beyond the Galaxy. Soon, several new experiments will continue to explore, with increased sensitivity, especially extended targets in the sky. This paper reviews the several near-term and longer-term plans for $\upgamma$-ray observatories, from MeV energies up to hundreds of TeV. Similarly, the X-ray sky has been and continues to be monitored by decade-old observatories. Upcoming telescopes will further bolster searches and allow new discovery space for lines from, e.g., sterile neutrinos and axion-photon conversion.

\begin{figure}
  \begin{center}
    \includegraphics[width=0.8\linewidth]{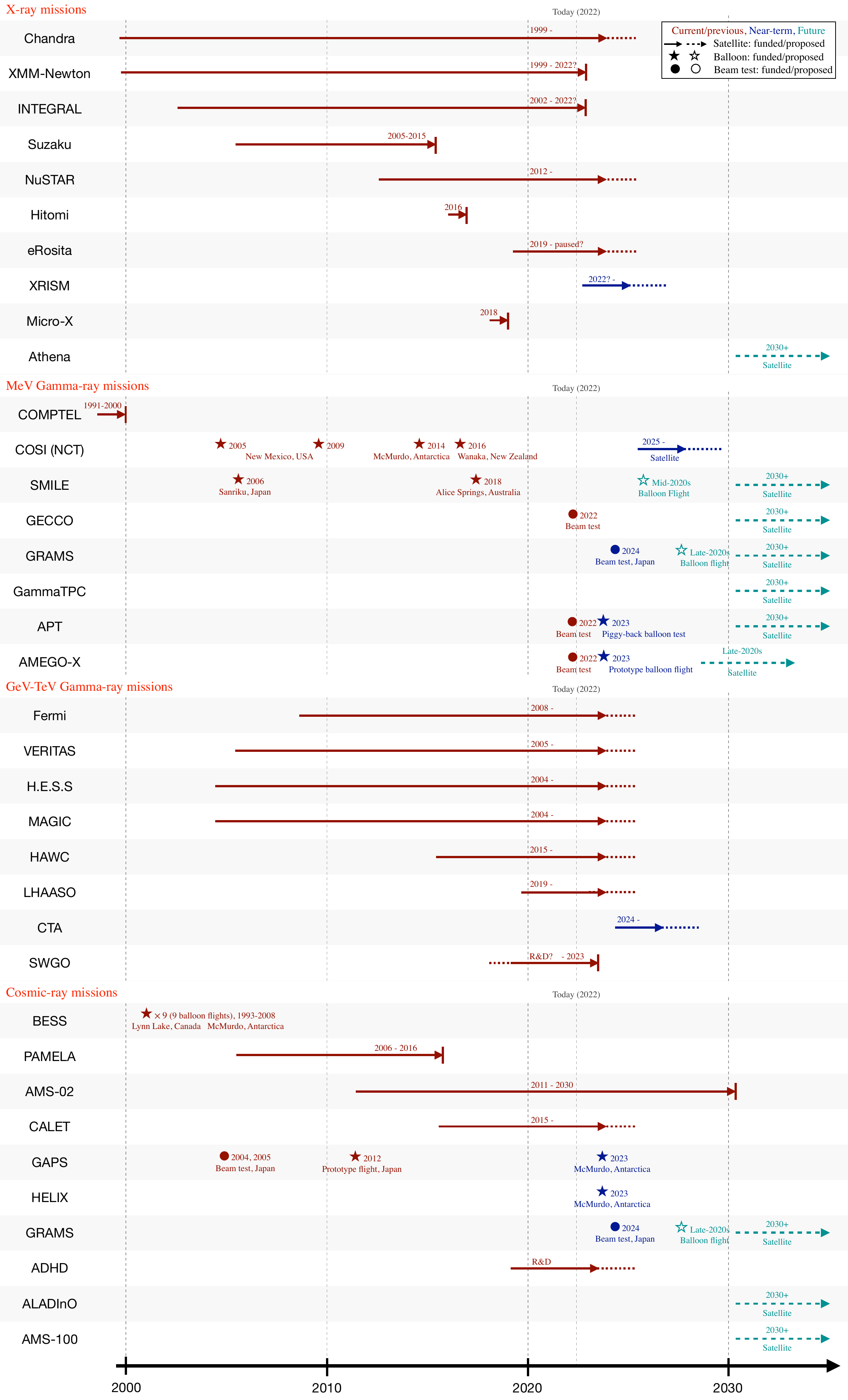}
    \end{center}
\caption{\label{fig:overview} {Overview of current, upcoming and proposed missions.}}
\end{figure}

Furthermore, this overview discusses currently operating cosmic-ray probes and the landscape of future experiments that will clarify existing persistent anomalies in cosmic radiation and spearhead possible new discoveries. Fig.~\ref{fig:overview} provides an overview of the different instruments.

Finally, the article closes with a discussion of necessary cross section measurements that need to be conducted at colliders to reduce substantial uncertainties in interpreting photon and cosmic-ray measurements in space.

\newpage

\tableofcontents

\newpage

\section{Introduction\label{s-intro}}

\textit{Contributors (alphabetical order): Esra Bulbul, Philip von Doetinchem, Kerstin Perez, Stefano Profumo, Brandon Roach}
\bigskip

The aim of this white paper is to give an overview of probes that provide sensitivity to particle dark matter by measuring cosmic high-energies photon and cosmic rays.  

This section provides a brief overview of classes of dark matter models that the current and future probes are sensitive to. The dark matter models can be structured in three broadly-defined categories: light, heavy, and super-heavy.  This classification is partly based upon mass ranges slated to produce signals relevant to different experimental facilities. Namely, ``heavy'' dark matter candidates refer to those with masses at or around the electroweak scale, such as weakly interacting massive particles (WIMPs~\cite{Arcadi:2017kky}). However, thermal production is not strictly assumed because what matters here is simply the energy scale at which annihilation or decay products are expected. ``Light'' dark matter models refer to those below the electroweak scale (GeV scale and below), and ``super-heavy'' dark matter models refer to those well above it (TeV scale and above).

Annihilation or decay of dark matter particles yields observable, stable particle species (e.g., electrons, photons, neutrinos) as a result of the fragmentation of prompt annihilation or decay products. Since dark matter in the late universe must predominantly be non-relativistic, the annihilation and decay products inherit the energy scale corresponding to the dark matter mass. Additionally, secondary radiation (e.g., synchrotron and bremsstrahlung off of electrons and positrons~\cite{Profumo:2010ya}) from the dark matter annihilation or decay products is predicted to populate the lower end of the electromagnetic spectrum, from soft gamma rays, to X-rays, to radio frequencies.

There are general, albeit model-dependent, upper limits on the mass of dark matter particles produced as thermal relics from the early universe~\cite{Griest:1989wd}. Assuming that the dark-sector particles are in thermal equilibrium with visible-sector particles, particles will, at least to some level, continue to annihilate in the late universe, leaving an observable imprint in the annihilation products and in the secondary radiation thereof (notice that there are model instances, such as with dominant coannihilation or models with a velocity-dependent cross section at low energies, where this might not be true). However, even if the dark matter particle is produced non-thermally, it can both annihilate or decay; in fact, this is in some cases theoretically predicted, such as in the case of dark matter whose stability is protected by a global symmetry~\cite{Mambrini:2015sia}.
 
\subsection{Super-heavy Dark Matter}

There are plentiful models for super-heavy dark matter (from above the electroweak scale up to the Planck scale): an incomplete model list includes WIMPzillas, strangelets, and Q-balls (see e.g.~\cite{Guepin:2021qai} and references therein)~\cite{SnowmassCF1WP8}. Non-thermal super-heavy particle creation was shown to be a generic phenomenon in the context of inflationary cosmology~\cite{Chung:1998zb}, and denoted ``{WIMPzillas!}'' in~\cite{Kolb:1998ki}. The flux of cosmic rays from the decay of a supermassive particle of a certain mass depends on the number density of particles along the line of sight, their decay rate, and the fragmentation function. Strangelets are nuggets of quarks that could form in a first-order phase transition (as first envisioned by Witten~\cite{ref329of1310.8642}) and be stable. These could be macroscopic clumps of quark matter with masses in the range $10^9-10^{18}$ g. Strangelets could be viable dark matter candidates (e.g.,~\cite{Carlson:2015daa}). Another interesting and a theoretically very well-motivated class of supermassive dark matter candidates is that of Q-balls (e.g.,~\cite{hepph0303065}). One of the phenomenological motivations for Q-balls is to explain baryogenesis and dark matter in one move and potentially connect that explanation to inflation. In all cases, cosmic-ray experiments, especially those with an extremely large effective area, have the best capabilities to detect decay debris off of super-heavy candidates.

\subsection{Heavy Dark Matter}

WIMP and WIMP-like particles' indirect detection is a long-standing and very well-estab\-lished field (see e.g. the recent review~\cite{Arcadi:2017kky}). 
The general idea is to observe the annihilation or decay of dark matter into standard-model particles as part of the $\upgamma$-ray and cosmic-ray spectra. As the most abundant cosmic-ray fluxes, like proton and helium, are dominated by production in standard astrophysical processes, like supernovae, the dark matter search focuses on finding distinct features in the spectra of less abundant species without dominant primary sources. An example for $\upgamma$-rays is the search for monochromatic line features~\cite{Bergstrom:1997fj}. For cosmic rays, the search concentrates on positrons and antinuclei from antiprotons to antihelium~\cite{Turner:1989kg,Aramaki:2015pii,vonDoetinchem:2020vbj,Carlson:2014ssa,SnowmassCF1WP6,SnowmassCF1WP7}. Dark matter searches with $\upgamma$-rays and cosmic rays require a precise understanding of the standard astrophysical background of a particular species. The ideal ``smoking gun'' signature is a cosmic messenger free of astrophysical background. However, this might come with the caveat of a low overall flux, leading to the need for large experiments with long measurement times.
New opportunities for WIMP-like particle indirect detection include the study of high-energy photons from the Galactic Center region with next-generation sensitivity~\cite{CTA:2020qlo}, observations of dwarf galaxies in the Southern Hemisphere for dark matter searches with a wide-field observatory~\cite{Viana:2019ucn}, and clarifying the origin of TeV Halos around pulsar wind nebulae as the origin of the cosmic-ray positron excess observed in the cosmic radiation~\cite{Profumo:2018fmz, Do:2020xli, SnowmassCF1WP6}. 
Furthermore, low-energy antideuterons and antihelium nuclei have been identified as a vital new signature of dark matter annihilation or decay, essentially free of astrophysical background~\cite{Coogan:2017pwt}. A first-time detection of low-energy cosmic antideuterons would be an unambiguous signal of new physics, opening a transformative new field of cosmic-ray research and probing a variety of dark matter models that evade or complement collider, direct, or other cosmic-ray searches (recent reviews: \cite{Aramaki:2015pii,vonDoetinchem:2020vbj}). 

\subsection{Light Dark Matter}

In the light dark matter realm, a renewed impetus surrounded the calculation of prompt MeV $\upgamma$-rays or X-rays from sub-GeV dark matter~\cite{Coogan:2019qpu}. 
Here, new facilities in space will allow the exploration of thus far completely unconstrained swaths of parameter space in the dark matter mass versus annihilation or decay rate, even though cosmology provides significant model-independent constraints on\,MeV dark matter models (e.g.,~\cite{Lehmann:2020lcv, DEramo:2018khz}); model-dependent phenomena, such as a dependence on the relative velocity of the annihilation rate, or production in celestial bodies~\cite{Leane:2021ihh,Leane:2021tjj}, also offer new opportunities for discovery.

Furthermore, astrophysical X-ray observations offer leading sensitivity to light dark matter candidates such as sterile neutrinos and axions. 
A sterile neutrino~\cite{PhysRevLett.72.17, Shi:1998km,Abazajian:2001nj,Asaka:2005an,Asaka:2006nq,Shaposhnikov:2006xi,Laine:2008pg,SnowmassCF1WP6} can decay, via its mixing with active neutrinos, into an active neutrino and a photon, providing a clear X-ray line signature. 
Axions~\cite{Peccei:1977hh,Peccei:1977ur} and axion-like particles (e.g.,~\cite{Jaeckel:2010ni, DiLuzio:2021ysg} appear in several extensions of the Standard Model of particle physics~\cite{Jaeckel:2010ni,Ringwald:2012hr} and are a generic prediction of string theory~\cite{Halverson2019}. 
Axions, via their enhanced axion-photon oscillations in the presence of strong electric or magnetic fields, can alter stellar processes and the propagation of light, visible as variations in the X-ray spectra of many astrophysical objects (see e.g.~\cite{Zioutas:2009bw}.

\section{Photon probes}\label{sec:photons}

\textit{Contributors (alphabetical order): Tsuguo Aramaki, James Buckley, Esra Bulbul, J. Patrick Harding, Chris Karwin, Jason Kumar, Rebecca Leane, Tom Melia, Kerstin Perez, Brandon Roach, Daniel Salazar-Gallegos, Julie McEnry, Miguel S\'anchez-Conde, Andrew W. Strong, Tom Shutt, Toru Tanimori, Atsushi Takada, John Tomsick, David Williams}
\bigskip

\subsection{GeV--TeV \texorpdfstring{$\upgamma$}{}-ray Experiments}

\subsubsection{Current Status}

\paragraph{Fermi Gamma-ray Space Telescope}

\begin{figure}
  \begin{center}
    \includegraphics[width=0.6\linewidth]{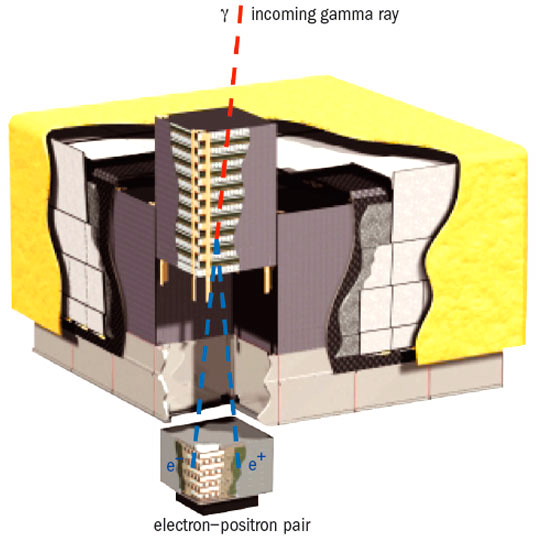}
    \end{center}
\caption{Overview of Fermi-LAT.\label{fermi}}
\end{figure}

The Fermi Gamma-ray Space Telescope was launched in 2008. The primary instrument (Fig.~\ref{fermi}) onboard is the Large Area Telescope (LAT), a pair conversion detector consisting of 16 tracker modules, 16 calorimeter modules, and a segmented anti-coincidence shield (ACS). An incident $\upgamma$-ray photon passes through the ACS and is converted into an electron-positron pair in the silicon-strip trackers, with the energy being subsequently deposited into the CsI calorimeters. These measurements allow for reconstructing the photon's incident energy and direction. The LAT is in low-Earth orbit (\til565 km) and operates primarily in survey mode, scanning the entire sky every two orbits (3.2\,h). It has a large field of view (2.4\,sr, or about one-fifth of the whole sky), a large effective area, and good energy and angular resolution. The energy sensitivity covers the range of 20\,MeV to \til1\,TeV. 

With over 13 years of data collection, many dark matter (DM) searches have been performed with the LAT. A systematic excess of $\gamma$ rays has been detected coming from the Galactic Center (GC) region~\cite{Goodenough:2009gk,Hooper:2010mq,abazajian2011,abazajian2014,calore2015background,Daylan2016,2015ApJ...812...15B,ajello2016fermi,TheFermi-LAT:2017vmf,DiMauro:2021raz,doi:10.1146/annurev-nucl-101916-123029,Bartels:2015aea,lee2016,leane2019dark,Zhong:2019ycb,Leane:2020nmi, Leane:2020pfc}. The source of this excess remains an open question, with the current leading explanations including mis-modeling of the Galactic diffuse emission along the line of sight, emission from a sub-threshold source population such as millisecond pulsars, or WIMP DM annihilation. To date, no complementary signal has been detected from a combined analysis of the Milky Way dwarf spheroidal satellite galaxies -- expected to be the cleanest objects for WIMP searches -- and numerous studies have placed upper limits on the DM annihilation cross section~\cite{2010ApJ...712..147A,GeringerSameth:2011iw,Ackermann:2013yva,Fermi-LAT:2015att,2015ApJ...809L...4D,Fermi-LAT:2016uux,2020PhRvD.101d3017L,2020PhRvD.102f1302A,2020JCAP...02..012H,DiMauro:2021raz}. These upper limits remain one of the most robust and stringent constraints from indirect DM searches and, specifically, they are crucial for DM interpretations of the GC excess. Other important studies have been conducted for numerous complementary targets, and they have provided competitive and independent upper limits as well, including those obtained from dwarf irregular galaxies~\cite{Gammaldi:2021zdm}, the Large and Small Magellanic Clouds~\cite{Buckley:2015doa,Caputo:2016ryl}, Galactic DM subhalos~\cite{BuckleyHooper10,fermi_dm_satellites_paper,Belikov2012,Zechlin+12,ZechlinHorns12,BerlinHooper14,Bertoni+15,Schoonenberg+16,Coronado-Blazquez:2019puc,2019JCAP...11..045C}, the Milky Way halo~\cite{Ackermann:2012rg,2013JCAP...10..029G}, M31~\cite{2019ApJ...880...95K,DiMauro:2019frs,Karwin:2020tjw}, galaxy clusters~\cite{2012JCAP...01..042H,2012JCAP...07..017A,2015ApJ...812..159A,2016JCAP...02..026A,2016PhRvD..93j3525L,2017NatSR...714895C,2018PhRvL.120j1101L}, the extragalactic $\upgamma$-ray background~\cite{2010JCAP...04..014A,2015JCAP...09..008F,Ajello:2015mfa} and DM signals towards the Sun~\cite{Mazziotta:2020foa}. Limits have also been placed on models of axion-like particles (ALPs), in this case looking for ALP-induced spectral distortions in LAT data~\cite{2016PhRvL.116p1101A,2016PhRvD..93d5019B,Kohri:2017ljt,Meyer:2020vzy,2021PhLB..82136611C,Crnogorcevic:2021wyj}.

\paragraph{Imaging Atmospheric Cherenkov Telescopes (IACTs)} 

Imaging atmospheric Cherenkov telescopes detect the Cherenkov flash produced in the atmosphere by relativistic secondary particles in the showers initiated by astrophysical $\upgamma$-rays and charged cosmic rays. Arrays of telescopes can capture images of the showers from multiple perspectives and use them to reconstruct the energy and direction of the primary $\upgamma$-ray and differentiate $\upgamma$-rays from the more numerous cosmic rays. Depending on the size and configuration of the telescopes, IACTs are sensitive to $\upgamma$-rays from \til20\,GeV to $>$30\,TeV. They achieve higher instantaneous sensitivity than extensive air shower detectors, such as HAWC, by virtue of lower energy threshold, better angular resolution, and better identification of the primary particle type. Conversely, they require clear, dark skies to operate and must be pointed at individual objects of interest, given their few-degree fields of view. Three major facilities of this type are in operation: VERITAS (the Very Energetic Radiation Imaging Telescope Array System in southern Arizona), H.E.S.S. (the High Energy Stereoscopic System in Namibia), and MAGIC (Major Atmospheric Gamma Imaging Cherenkov Telescopes in Spain on the Canary Islands). They have been in operation for 15--20 years and have long observations of the most promising targets for dark matter annihilation or decay. The most constraining dark matter limits typically come from observations of the Galactic Center region~\cite{HESS:2006zwn, HESS:2011zpk, HESS:2013rld, HESS:2016mib}, which is complicated because of the presence of astrophysical sources of $\upgamma$-rays but is the closest large collection of dark matter. Dark matter limits have also been reported from collections of dwarf galaxies~\cite{VERITAS:2010meb,HESS:2014zqa,MAGIC:2016xys,VERITAS:2017tif,HESS:2018kom,HESS:2020zwn}, galaxy clusters~\cite{VERITAS:2012bez}, and candidate dark matter clumps within the Galaxy~\cite{Nieto:2015hca}.

\paragraph{High Altitude Water Cherenkov Observatory (HAWC)} 

The High-Altitude Water Cherenkov Observatory (HAWC) is a water Cherenkov detector array located at Sierra Negra, Mexico. HAWC is an extensive air shower detector that detects charged particles in particle showers generated by high-energy $\upgamma$-ray or cosmic-ray interactions in the atmosphere~\cite{HAWC_CRAB,HAWC:2019xhp,HAWC:2020hrt}. HAWC has a near 100\% duty cycle with a 2\,sr field of view and a 22,000\,m$^2$ effective area. This complements IACTs with their smaller fields-of-view. HAWC’s $\upgamma$-ray energy sensitivity ranges from \til300\,GeV to 100s of TeV. Cosmic bodies that pass within HAWC’s sights include the Galactic Plane, Virgo cluster, Andromeda (M31), Crab nebula, high energy blazars Markarians 421 and 501.

HAWC is best for dark matter searches from extended sources like M31~\cite{HAWC:2018eaa} or our local Milky Way dark matter halo~\cite{HAWC:2017udy}, in part from its wide field of view. HAWC dark matter searches towards extended sources have produced competitive limits for the field in the TeV energy range. HAWC searches for dark matter events from the Sun~\cite{HAWC:2018szf} as it is very close and occludes $\upgamma$-rays at the TeV scale while producing very few -- IACTs cannot match HAWC's solar searches with this clarity. Data taken with HAWC is archived each day, so newly suspected dark matter clumps can be searched in the full HAWC dataset~\cite{HAWC:2018etk}, including observations towards dwarf spheroidal galaxies~\cite{HAWC:2017mfa,HAWC:2019jvm}. HAWC can also look for transient sources, opening searches for primordial black holes~\cite{HAWC:2019wla}, or coincidental searches with other observatories~\cite{Hess:2021cdp}. Though HAWC has not found positive dark matter signals, the limits published by the collaboration are competitive at the TeV-PeV scale of dark matter mass. 

\paragraph{Large High Altitude Air Shower Observatory (LHAASO)} 

LHAASO is located in the Northern hemisphere in the Sichuan province, China, and started operation in 2019. It follows the HAWC and SWGO water Cherenkov design with an effective area of 78000 m$^2$ along with a high-energy muon detector array of 1.3\,km$^2$ area~\cite{10.1088/1674-1137/ac3fa6}. 
With its large field of view, it is sensitive to $\upgamma$-rays at tens of TeV.

LHAASO will conduct dark matter searches in various astrophysical sources. However, it needs to be noted that due to its northern-hemisphere location, LHAASO will have difficulty observing the Galactic Center (the Galactic Center peaks at 58$^\circ$ from zenith for LHAASO). Therefore, LHAASO dark matter studies will not easily leverage the large nearby dark matter halo from the Galactic Center for their searches. LHAASO will, however, have strong sensitivity to dark matter events from the Sun, potentially exceeding the HAWC sensitivity~\cite{Leane:2017vag}. For more information about the LHAASO program on beyond-the-standard-model physics, see~\cite{10.1088/1674-1137/ac3fab}.

\subsubsection{Near-term Future}

\paragraph{Cherenkov Telescope Array (CTA)} 

The Cherenkov Telescope Array Observatory (CTAO) will be a next-generation IACT facility to study astrophysical sources of $\upgamma$-rays from 20\,GeV to 300\,TeV~\cite{2011ExA....32..193A,
CTAConsortium:2013ofs,2019scta.book.....C,2019APh...111...35A}.  It is designed to have improved sensitivity by a factor between five and twenty (depending on the energy) compared to the current generation IACTs. To study the entire sky, there will be telescope installations in both the Northern and Southern Hemispheres. In the north, telescopes will be in Spain on La Palma, one of the Canary Islands. The southern installation will be in the Atacama desert in Chile, within the grounds of the European Southern Observatory.  

One of the principal science motivations for CTA is the search for dark matter, and it has been studied extensively by the CTA Consortium~\cite{2019scta.book.....C, CTA:2020qlo}. 
The unparalleled 20\,GeV--300\,TeV sensitivity of CTA enables indirect dark matter searches by observing cosmic targets where a WIMP annihilation signal may be discernible from other astrophysical processes. 
Given its relative proximity and expected large dark matter density, the Galactic Center region will be an essential target for this purpose. 
For the canonical velocity-averaged thermal annihilation cross section of 
$\til3 \times 10^{-26}$ cm$^3$ s$^{-1}$, CTA will be able to detect annihilation into several of the expected channels for a WIMP mass in the range \til0.2--20\,TeV \cite{CTA:2020qlo}, something which is not possible at the higher masses with current instruments {\it of any type}. Together with Fermi-LAT constraints on dark matter lighter than \til200 GeV \cite{Fermi-LAT:2015att}, the WIMP phase space will be severely constrained in the case of non-detection.
Other CTA studies, such as TeV halo observations around nearby pulsar wind nebulae~\cite{HAWC:2017kbo} and cosmic ray electron-positron spectrum measurements up to hundreds of TeV~\cite{2019scta.book.....C,ctaelectrons}, will also be fundamental to understanding potential signatures of WIMP annihilation or decay in cosmic ray indirect DM searches.

An application to the European Union is in preparation to form a European Research Infrastructure Consortium (ERIC) for the construction and operation of CTAO. Funding has been identified and committed to building an ``Alpha Configuration'' of the observatory during 2022--2027. 
The Alpha Configuration will have capabilities greatly exceeding any of the existing IACT arrays while being constrained by available funding to have fewer telescopes than initially envisioned for CTA. In particular, the southern observatory will have 14 ``medium-sized telescopes,'' which are sensitive in the core energy range of 100\,GeV--10\,TeV, compared to a goal of 25. 

An international consortium of CTA members, led by the U.S., has developed and prototyped a novel medium-sized telescope design for CTA, called the Schwarzschild-Couder Telescope (SCT), incorporating a secondary mirror to substantially improve the performance 
~\cite{2007APh....28...10V,2008ICRC....3.1445V, Adams:2020qxs}. The addition of \til10 SCTs to CTAO in the south, with lead mid-scale funding from NSF in collaboration with other agencies, domestic and abroad, would bring the performance of CTA for dark matter studies of the Galactic Center region at least to the originally planned level anticipated in the studies above. In addition to augmenting the telescope count, the SCT has much improved angular resolution over the full 8$^\circ$ field of view, compared to the telescopes in the Alpha Configuration. A dark matter signal from the Galactic Center halo is a moderately diffuse source superimposed on more localized astrophysical signals. The superior angular resolution of the SCT over a wide field of view will bring added power to separate astrophysical signals from any produced by dark matter.  

\paragraph{Southern Wide-Field Gamma-ray Observatory (SWGO)} 

The Southern Wide-field Gamma-ray Observatory (SWGO)~\cite{swgo_instrument,swgo_science} is a water Cherenkov detector array and planned to be located in the Southern Hemisphere, having a sensitivity \til10$\times$ better than the High-Altitude Water Cherenkov (HAWC) Observatory~\cite{HAWC_CRAB}. Both measure relativistic particles in extensive air showers caused by cosmic-ray and $\upgamma$-ray interactions in the atmosphere. These arrays have a wide field-of-view and observe \til2/3 of the sky every day with a near-100\% duty cycle. They complement Imaging Atmospheric Cherenkov Telescopes (IACTs), which have smaller fields-of-view. For example, SWGO will observe extended objects, like the regions relatively far from the GC, allowing for backgrounds that minimize contamination from $\upgamma$-ray sources, thus increasing its sensitivity to emission from the wider dark matter halo.

With its wide field-of-view and TeV-energy sensitivity, SWGO will search for dark matter in various astrophysical sources: galaxy clusters, dwarf galaxies, the Andromeda galaxy, the Magellanic clouds, the Sun, the diffuse emission from the Milky Way, and the Milky Way Galactic Center. The searches for dark matter in the Galactic Center will be of particular interest for SWGO. The ability to observe a more extended region makes the SWGO sensitivity less dependent on the assumed behavior of the dark matter density profile than pointed IACTs.
Hence, SWGO will provide robust dark matter limits from the Galactic Center with different systematic uncertainties from those obtained with the narrower fields of view of IACTs.
Furthermore, because SWGO can look at a large area of the sky covered by Galactic dark matter, it can look at more than ten times as much Galactic dark matter as targeted searches in the Galactic Center itself. SWGO will also look at the whole sky during its observations, enabling joint searches for dark matter from dwarf galaxies. The Rubin Observatory~\cite{Mandelbaum:2018ouv} will survey the Southern Hemisphere sky with unprecedented sensitivity and is expected to find hundreds of new dwarf spheroidals ~\cite{Hargis:2014kaa}. Legacy data from SWGO at these locations could easily and immediately be analyzed when new dwarf spheroidals are found. SWGO will also have strong sensitivity to solar gamma rays, and may therefore provide a solar dark matter probe stronger than both HAWC and LHAASO in the TeV $\upgamma$-ray range~\cite{Nisa:2019mpb}.

\subsection{MeV \texorpdfstring{$\upgamma$}{}-ray experiments}

\subsubsection{Current Status}

\paragraph{Imaging Compton Telescope (COMPTEL)}

\begin{figure}
  \begin{center}
    \includegraphics[width=0.9\linewidth]{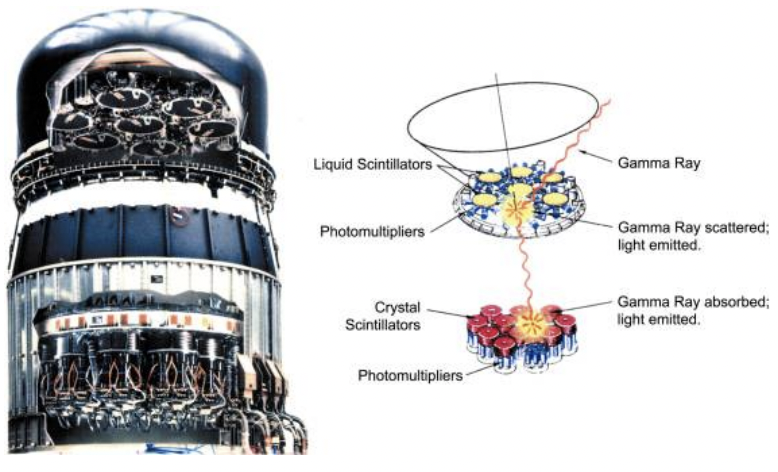}
    \end{center}
\caption{\label{fig:COMPTEL} {The COMPTEL instrument and principle of operation.}}
\end{figure}

COMPTEL was the Compton telescope on NASA’s Compton $\upgamma$-ray Observatory (CGRO) launched in 1991 and which was re-entered in 2000~\cite{schonfelder1993instrument, strong2019comptel}. COMPTEL covered the energy range 0.75 to 30\,MeV, and performed a full-sky survey. The main achievements of the COMPTEL mission included Galactic and extragalactic sources~\cite{schonfelder2000first}, $^{26}$Al maps, GRBs, solar flares, and the extragalactic diffuse $\upgamma$-ray background.

COMPTEL was a double-scatter Compton telescope: incoming $\upgamma$-rays Compton-scatter in one of the seven upper organic liquid-scintillator (D1) detectors, and are absorbed in one of the 14 lower NaI (D2) detectors (see Fig.~\ref{fig:COMPTEL}). Both D1 and D2 use photomultipliers to measure the light signal and locate the scatter position using the Anger-camera principle. The energy deposits give the Compton scatter angle. Hence the incoming direction is determined to an annulus on the sky, whose width depends on the precision of the energy and position measurements. At high energies, the absorption in D2 is incomplete, so the response is correspondingly broadened. The angular resolution of the Compton scatter angle is about 2$^{\circ}$. The distance between D1 and D2 is 1.577\,m, allowing a time-of-flight (TOF) discrimination for upward-moving background $\upgamma$-rays. A plastic-scintillator anticoincidence dome surrounding the instrument reduces the charged-particle background. In addition, a pulse-shape-discrimination (PSD) measurement is used for background rejection. Nevertheless, the data are background-dominated, which necessitates suitable background-handling methods. In its 9.7\,years of operation, COMPTEL performed about 340 pointings each of roughly two weeks duration with field-of-view radius about 30$^{\circ}$, covering the entire sky. 

\subsubsection{Near-term Future}

\paragraph{Compton Spectrometer and Imager (COSI)}

\begin{figure}
  \begin{center}
    \includegraphics[width=0.8\linewidth]{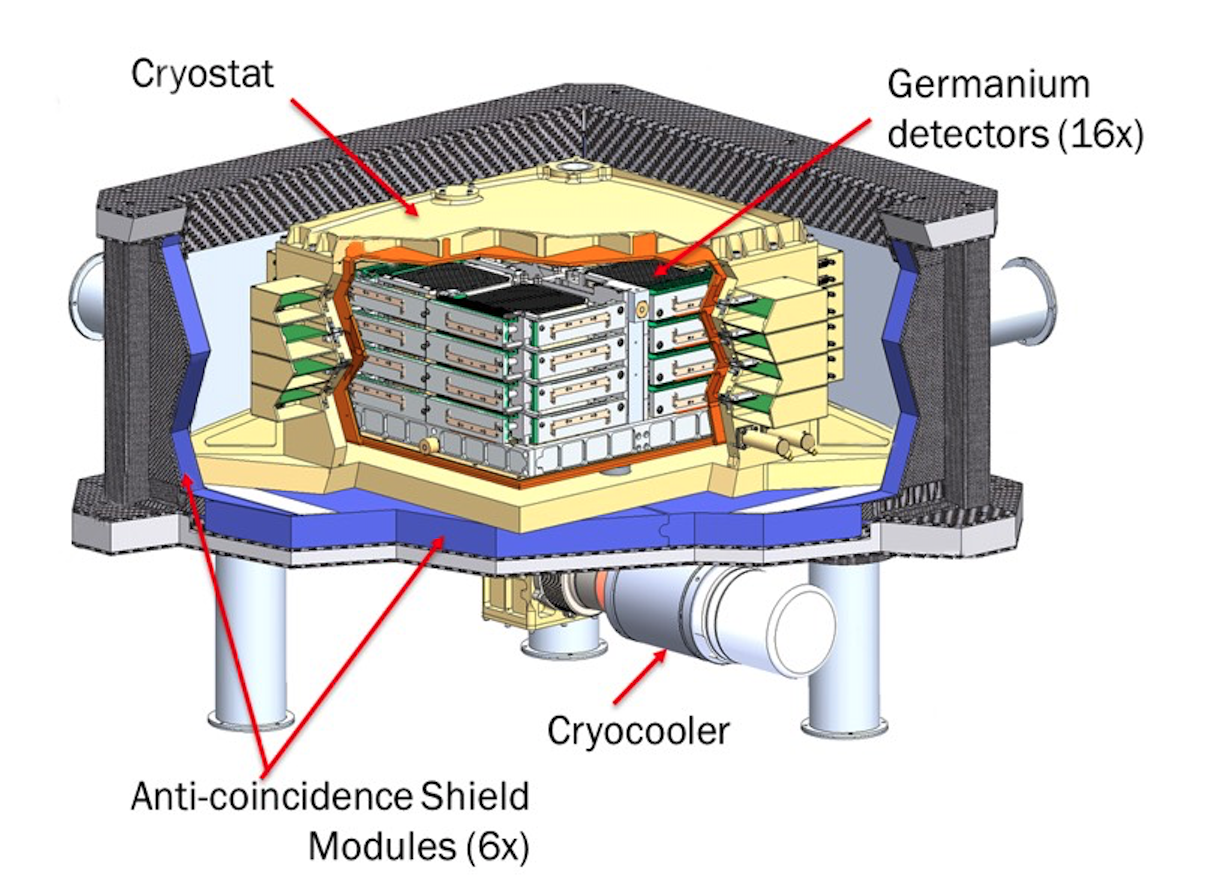}
    \end{center}
\caption{\label{fig:COSI} {Cutaway view of the COSI instrument.}}
\end{figure}

The Compton Spectrometer and Imager (COSI) is a wide-field telescope designed to survey the $\upgamma$-ray sky at 0.2--5\, MeV~\cite{Tomsick:2021wed}. COSI has been selected as a NASA Small Explorer (SMEX) satellite mission with a planned launch in 2025. Like the previous COMPTEL mission, COSI operates as a Compton telescope but with major advances in capabilities. COSI is designed to have a very large field of view (FOV), covering $>$25\% of the sky instantaneously and the full sky every day. The large FOV is combined with excellent energy resolution ($<$1\% FWHM), allowing for Galaxy-wide measurements of emission lines, including the electron-positron annihilation line at 0.511\,MeV and nuclear lines at 1.157, 1.173, 1.333, and 1.809\,MeV. In addition to imaging and spectroscopy, COSI will be capable of measuring the polarization of astrophysical sources such as $\upgamma$-ray bursts (GRBs) and accreting black holes.

COSI employs a novel design using a compact array of cross-strip germanium detectors (GeDs) to resolve individual $\upgamma$-ray interactions in the GeDs with high spectral and 3-dimensional spatial resolution, making COSI operate as a Compton telescope (see Fig.~\ref{fig:COSI}). The COSI array of 16 GeDs is housed in a common vacuum cryostat cooled by a mechanical cryocooler. The GeDs are read out by custom ASIC electronics integrated into the data acquisition system. An active bismuth germanate (BGO) shield encloses the cryostat on the sides and bottom to veto events outside the FOV. 

Annihilating or decaying light dark matter produces significant photon emission, including in general an X-ray and\,MeV $\upgamma$-ray continuum arising directly from the final state~\cite{Essig:2013goa} as well as from inverse Compton scattering on CMB and stellar radiation by final state electrons and positrons~\cite{Cirelli:2020bpc}. Due to its combination of sensitivity and sky coverage, COSI will provide a significant tightening over previous constraints on decaying or annihilating light DM. The sensitivity of COSI to the annihilation rate of light dark matter into a pair of photons can be estimated in the following way: Taking the annihilation signal from a 10$\times$10$^\circ$ region about the Galactic Center and assuming an Einasto dark matter profile, COSI is sensitive to the velocity-averaged annihilation cross sections of $3.5\times10^{-35}\,$cm$^3$/s, $1.4\times10^{-34}\,$cm$^3$/s, and $1.3\times10^{-33}\,$cm$^3$/s, for dark matter masses 0.3\,MeV, 1\,MeV, and 3\,MeV, respectively. The quoted numbers are the five sigma discovery reach, assuming a line search from 0.2\,MeV to 5\,MeV in 0.005\,MeV bins.

\subsubsection{Proposed Future Missions\label{s-mevvision}}

\paragraph{Sub-MeV $\upgamma$-ray Imaging Loaded-on balloon Experiment (SMILE)}

\begin{figure}
  \begin{center}
    \includegraphics[width=1\linewidth]{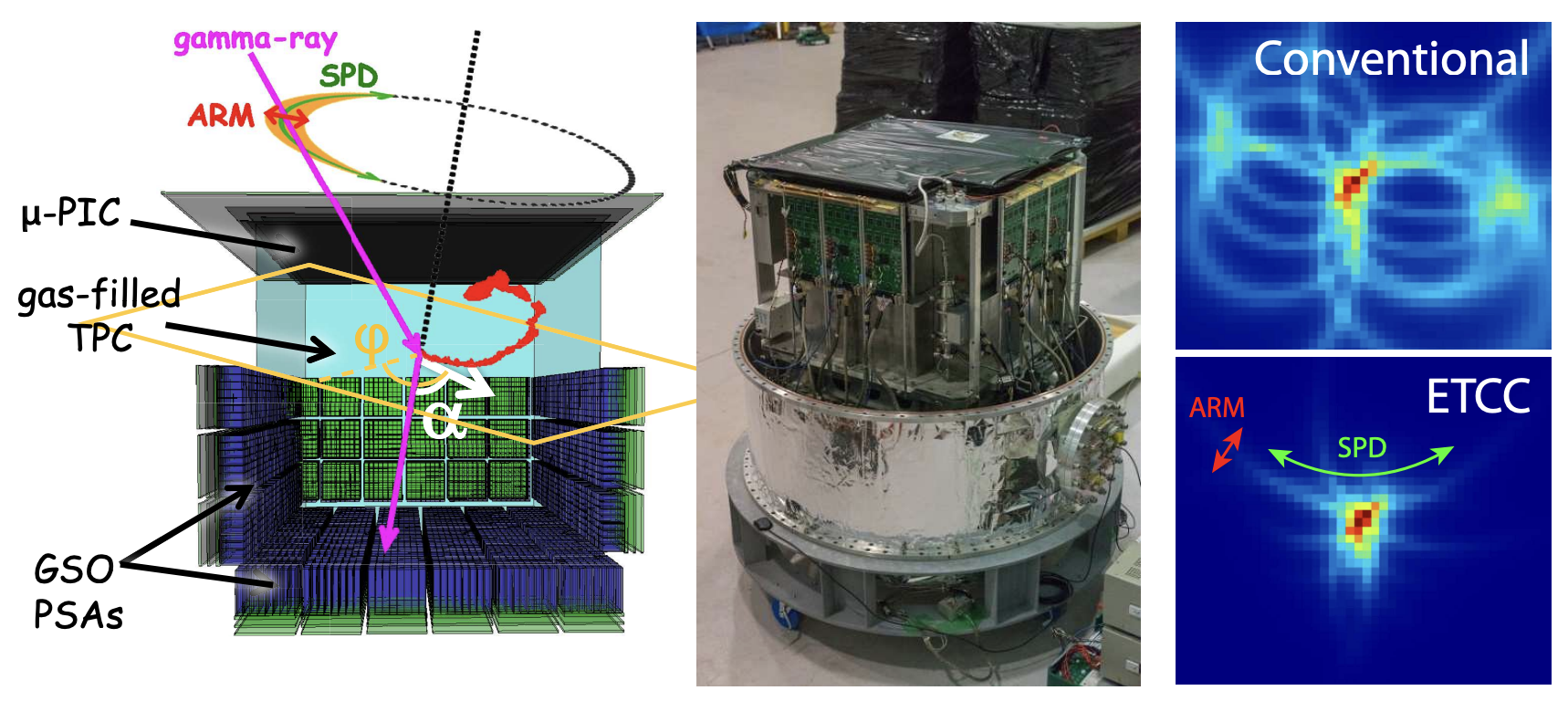}
    \end{center}
\caption{\label{fig:Smile} {(Left) Schematic view of SMILE-2+ 30\,cm-cubic ETCC. (Middle) Photograph of SMILE-2+ flight model instrument (Right) Point source images with the conventional Compton camera (top) and with SMILE2+ ETCC (bottom)}}
\end{figure}

The SMILE team has developed an Electron Tracking Compton Camera (ETCC) that can determine the direction of a $\upgamma$-ray as an arc while providing a bijection/linear image for the first time (Fig.~\ref{fig:Smile}), similar to other telescopes for X-rays and GeV $\upgamma$-rays. For this aim, a gaseous Time Projection Chamber (TPC) was used both as a scatterer and an electron tracking device. In 2018, SMILE observed the Galactic Center and detected the Crab, Galactic diffuse MeV $\upgamma$-rays, and cosmic diffuse MeV $\upgamma$-rays with the ETCC\cite{takada2021first,tanimori2020mev,hamaguchi2019space}. The SMILE collaboration is currently moving forward with the SMILE-3 experiment that can potentially have five times better sensitivity than COMPTEL with a one-month duration balloon flight. The proposed SMILE-3 mission can provide precise spectra of Galactic and cosmic diffuse MeV $\upgamma$-rays as well as source distribution of $\upgamma$-ray lines for 511\,keV, $^{26}$Al, $^{60}$Fe in the Galaxy~\cite{takada2020smile}.

Using a gas detector for MeV $\upgamma$-ray measurements is the core of the SMILE project. Except for pinhole cameras, conventional Compton cameras can only provide a non-linear image with the direction of a $\upgamma$-ray being a circle on the sky, which may often misidentify the source direction. Therefore, considering the high background in the MeV region produced by cosmic rays, the ETCC bijection/linear imaging system can have a huge advantage for obtaining the inverse mapping without degrading the sensitivity~\cite{tanimori2017establishment,komura2017imaging}. The SMILE ETCC consists of a gas TPC for tracking recoil electrons and GSO Pixel Scintillator Arrays (PSA) (Fig. \ref{fig:Smile}). The ETCC measures all parameters of the Compton kinematic equation, including the energy loss rate ($\text{d}E/\text{d}x$) of a recoil electron and the event topology, while providing a Point Spread Function (PSF) to identify the direction of the incident $\upgamma$-rays with a linear imaging system [150, 151]. The SMILE-3 ETCC, consisting of four 50-cm-cubes filled with 3\,atm CF4 gas, can achieve 2$^\circ$ of PSF (Scatter Plane Deviation (SPD) = 10$^\circ$ and Angular Resolution Measure (ARM) = 5$^\circ$), providing a sensitivity of <1 m-Crab with a 200\,cm$^2$ effective area in 10$^6$\,s observation time \cite{komura2017imaging}.

\paragraph{Galactic Explorer with a Coded Aperture Mask Compton Telescope (GECCO)}

GECCO is a novel concept for a next-generation $\upgamma$-ray telescope that will cover the hard X-ray$-$soft to $\upgamma$-ray region, and is currently being considered for a future NASA Explorer mission~\cite{moiseevsnowmass2021}. GECCO will conduct high-sensitivity measurements of the cosmic $\upgamma$-ray radiation in the energy range from 100\,keV to \til10\,MeV and create intensity maps with high spectral and spatial resolution, with a focus on the separation of diffuse and point-source components. These observations can help disentangle astrophysical and dark matter explanations of emission from the Galactic Center, and potentially provide a key to discovering as-of-yet unexplored dark matter candidates. 

\begin{figure}
  \begin{center}
    \includegraphics[width=1\linewidth]{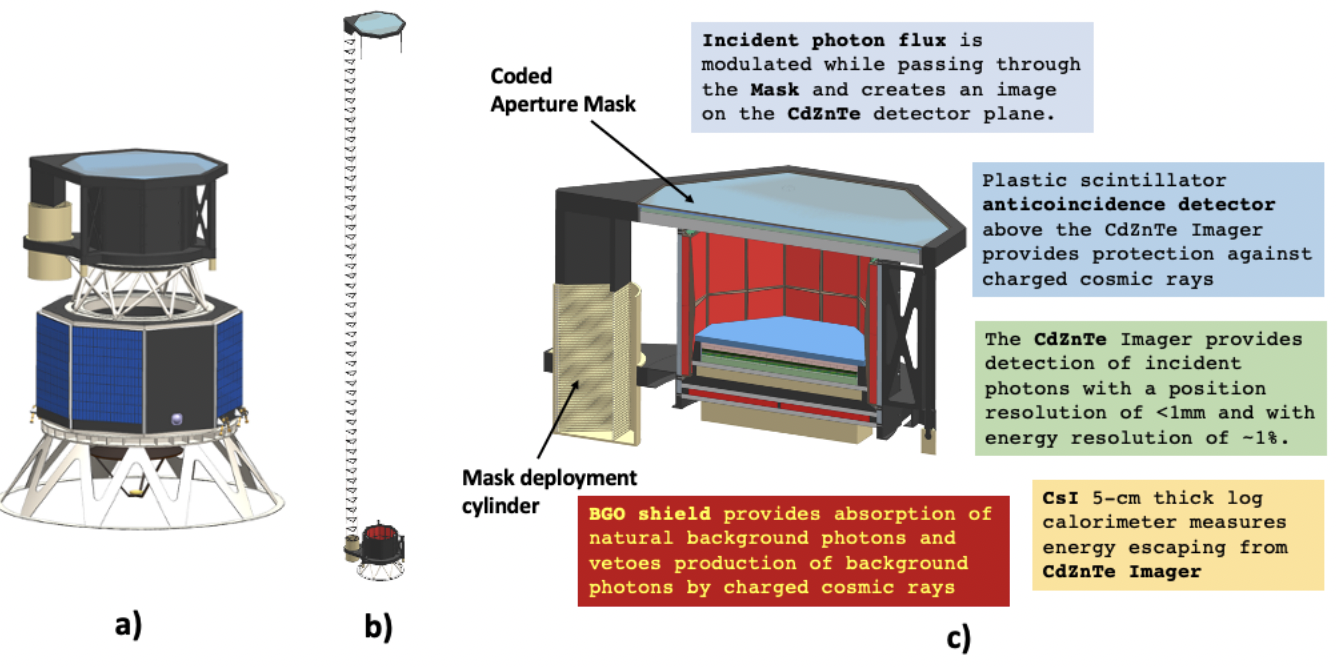}
    \end{center}
\caption{\label{fig:GECCO} {GECCO conceptual design: a) GECCO with Mask in stowed position and notional spacecraft bus, b) GECCO with Mask in deployed position, c) GECCO, cutaway}}
\end{figure}

The instrument (Fig.~\ref{fig:GECCO}) is based on a novel CdZnTe Imaging calorimeter and a deployable coded aperture mask. It utilizes a heavy-scintillator shield and plastic scintillator anti-coincidence detectors. The unique feature of GECCO is that it combines the advantages of two techniques -- the high-angular resolution possible with coded mask imaging, and a Compton telescope mode providing high sensitivity measurements of diffuse radiation. With this combined ``Mask+Compton'' operation GECCO will separate diffuse and point-sources components in the GC radiation with high sensitivity. GECCO will be operating mainly in pointing mode, focusing on the Galactic Center and other regions of interest. It can be quickly re-pointed to any other region, when alarmed. The expected GECCO performance is as follows: energy resolution $<$ 1\% at 0.5$-$5\,MeV, angular resolution \til1\,arcmin in Mask mode (5--6$^{\circ}$ field-of-view, \til2000\,cm$^2$ effective area), and 3$-$5$^{\circ}$ in the Compton mode (15$-$20$^{\circ}$ field-of-view, \til500\,cm$^2$ effective area). The sensitivity is expected to be better than 10$^{-6}$\,MeV/cm$^2$/s at 1\,MeV. These parameters are particularly promising for searching for dark matter
particles with\,MeV-scale masses and primordial black holes with 10$^{17}$\,g masses~\cite{Carr:2009jm,Carr:2016hva,Carr:2020gox,Ray:2021mxu,Speckhard:2015eva}.

\paragraph{Gamma-ray and AntiMatter Survey (GRAMS) }

\begin{figure}
  \begin{center}
    \includegraphics[width=1\linewidth]{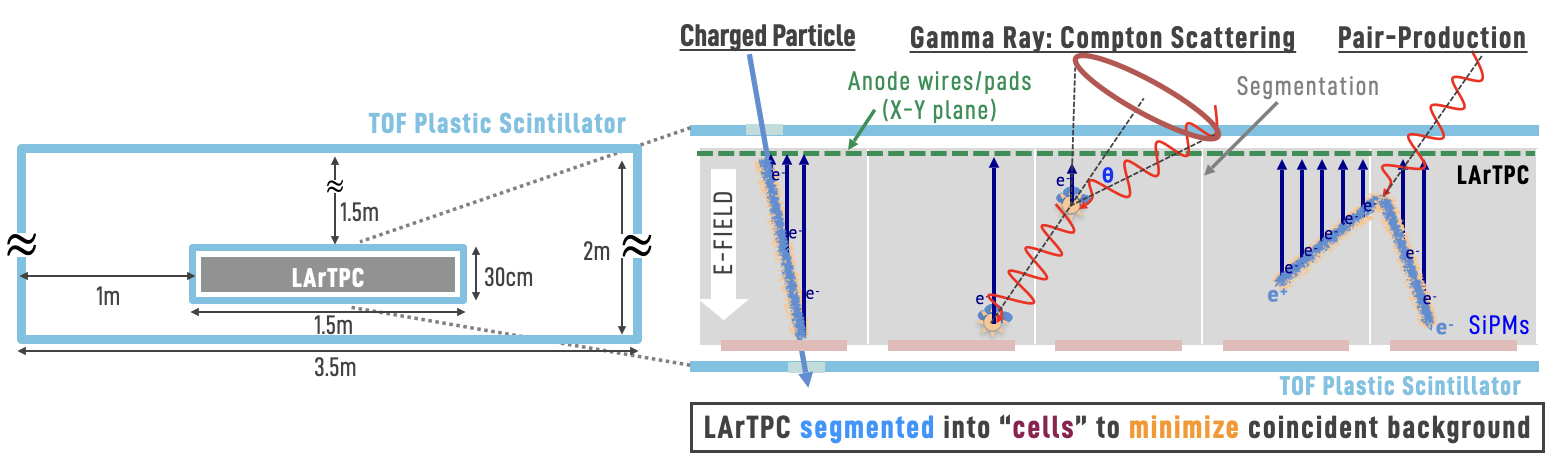}
    \end{center}
\caption{\label{fig:GRAMS} {GRAMS instrumental design.}}
\end{figure}

The Gamma-ray and AntiMatter Survey (GRAMS) project is a proposed next-generation balloon or satellite mission that simultaneously targets both astrophysical observations with MeV $\upgamma$-rays and indirect dark matter searches with antimatter~\cite{Aramaki:2019bpi, aramaki2020snowmass}. The MeV $\upgamma$-ray energy range has long been under-explored due to the lack of large-scale detectors to reconstruct Compton scattering events efficiently. GRAMS aims to break through existing technological barriers while utilizing a cost-effective, large-scale Liquid Argon Time Projection Chamber (LArTPC) detector as a Compton camera (Fig.~\ref{fig:GRAMS}). The LArTPC technology, successfully developed for underground dark matter and neutrino experiments over the last two decades, provides 3-dimensional particle tracking capability by measuring ionization charge and scintillation light produced by particles entering or created in the argon medium. 

GRAMS will provide an affordable, scalable, and full-sky-reach solution for a Compton telescope concept with the LArTPC. The GRAMS instrumental design includes a large-scale (1.4\,m$\times $1.4\,m$\times$20\,cm) LArTPC surrounded by two layers of plastic scintillators. The plastic scintillators veto charged particles while the LArTPC works as a Compton camera. The LArTPC volume will be segmented into small cells, localizing the signal and minimizing the coincident background events. GRAMS will provide an order of magnitude improved sensitivity to MeV $\upgamma$-rays with a single long-duration balloon flight. The GRAMS satellite mission could provide another order of magnitude improved sensitivity. Note that the GRAMS detector configuration also offers extensive sensitivities to antinuclei potentially produced by dark matter particles (Sec.~\ref{sec:GRAMS}).

\paragraph{GammaTPC}

\begin{figure}
  \begin{center}
    \includegraphics[width=0.8\linewidth]{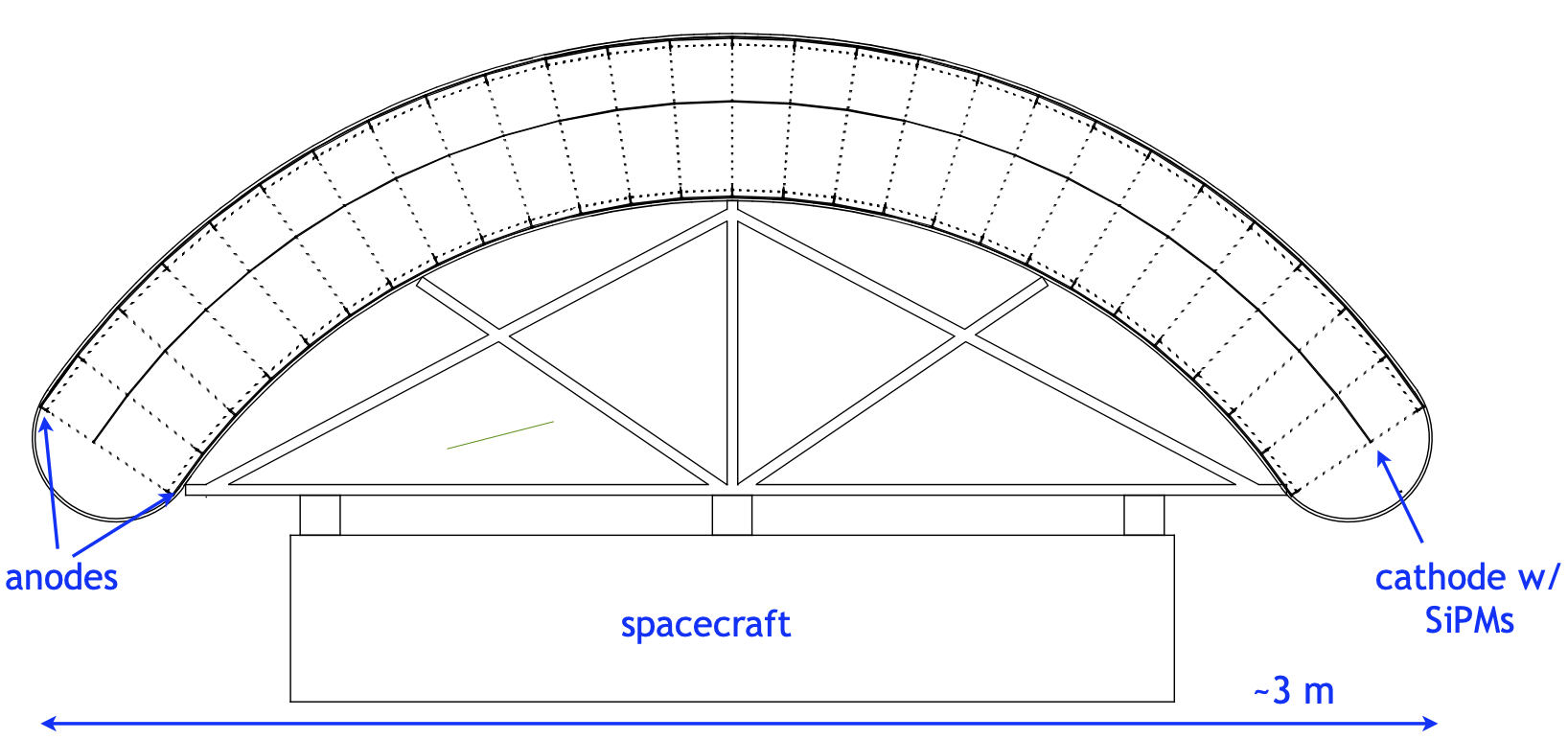}
    \end{center}
\caption{\label{fig:GammaTPC} {Instrumental Design of GammaTPC}}
\end{figure}

The goal of GammaTPC is to develop a 0.1--10\,MeV $\upgamma$-ray satellite instrument with a very large effective area and large field of view based on LArTPC technology (Fig.~\ref{fig:GammaTPC}). As mentioned above for GRAMS, an LArTPC promises a large instrument with modest channel count and power consumption, both of which are highly constrained in space. The program aims to develop needed technology to maximally exploit the LArTPC technology, and give pointing and energy resolution at least equivalent to Si strip based instruments (e.g., AMEGO-X), which GammaTPC is a natural successor to. In the current era of sharply reduced cost to launch large mass in space, The GammaTPC collaboration is targeting a \til10\,m$^2$, \til4\,tonne class instrument for a MIDEX mission. Such an instrument would have about two orders of magnitude larger effective area than currently proposed missions, and would be a very significant advance in science reach in this little-explored energy range.

The technology rests on significant developments in recent years in liquid noble TPCs for dark matter and neutrino physics, and in particular the DUNE program. The detector will feature a highly segmented set of TPCs housed in carbon fiber shell, with light readout using SiPMs and waveshifter. The segmentation is necessary to cope with the large flux of particles in low Earth orbit. A core new development is a novel Dual Scale Charge Readout (DSCR) that provides fine grained (\til500\,$\upmu$m pitch) pixel readout. The pixels are power switched based on signals from a cm-scale coarse grid that also is used to obtain a charge integral without loss of charge to sub-threshold sensors. This scheme drastically reduces power consumption, enabling high resolution readout. Several other developments needed for the space environment can be economically demonstrated with CubeSats.

\paragraph{Advanced Particle-astrophysics Telescope (APT) }

\begin{figure}
  \begin{center}
    \includegraphics[width=0.6\linewidth]{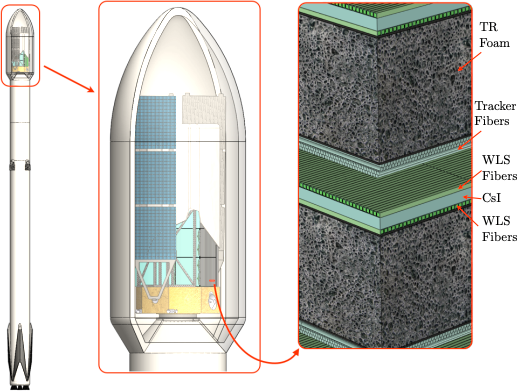}
    \end{center}
\caption{\label{APT}{APT in Falcon-9 faring.}}
\end{figure}

The Advanced Particle-astrophysics Telescope (APT) (Fig.~\ref{APT}) is a concept for a future probe-class space-based $\upgamma$-ray and cosmic-ray detector~\cite{buckleyastro2020}. The mission would have a broad impact on astroparticle physics. However, the primary science drivers for the mission include (1) probing WIMP dark matter across the entire natural mass range and annihilation cross section for a thermal WIMP, (2) providing a nearly all-sky instantaneous FoV, with prompt sub-degree localization and polarization measurements for $\upgamma$-ray transients such as neutron-star mergers and (3) making measurements of rare ultra-heavy cosmic-ray nuclei to distinguish between n-star merger and SNae r-process synthesis of the heavy elements. A first prototype of the new detector elements was successfully flown on a piggy-back Antarctic flight. The ADAPT (Antarctic Demonstrator for APT) suborbital mission was recently funded by NASA and is planned for a 2025 Antarctic flight. 

APT will combine a pair tracker and Compton telescope in a single monolithic design. By using scintillating fibers for the tracker and wavelength-shifting fibers to readout CsI detectors, the instrument will achieve an order of magnitude improvement in sensitivity compared with Fermi but with fewer readout channels and lower complexity. Advances in scintillating fibers and solid-state photomultipliers are critical enabling technologies for the approach. Another key technical advance is the development of high-speed waveform capture electronics that can capture the prompt signals from plastic scintillators and the slow scintillation signal from sodium doped CsI. 
The APT instrument consists of a distributed imaging calorimeter of limited depth (about 5.5\,radiation length) comprised of 20\,layers of thin (5\,mm) CsI:Na tiles readout by crossed WLS fibers (for imaging) and SiPM edge detectors (to improve calorimetry) (Fig.~\ref{APT}). The instrument can achieve adequate energy resolution (better than 30\% up to TeV energies) to provide good event reconstruction by fitting the shower profile. The instrument has 20\,layers of interleaved tracking fibers, with each $x$ or $y$-plane made of two interleaved layers of 1.5\,mm round fibers. 
Since there are no passive converter layers, the instrument provides a dramatic improvement in sensitivity at 10-100 \,MeV compared to Fermi and an energy resolution reaching 10\%, well below current or proposed pair telescopes. In addition, the ability to fit the profile of the shower development allows adequate energy reconstruction up to TeV energies. This leads to about 19\,times the sensitivity of Fermi in the higher energy regime (100\,MeV to 1\,TeV). The instrument is planned to be launched by a heavy lift vehicle into a Lagrange orbit to minimize Earth obscuration. The large effective geometric area (3\,m$\times$3\,m) and symmetry for upward and downward-going events will provide an effective area of \til20\,m$^2$sr at 1\,MeV and 1\,GeV.

\paragraph{All-sky Medium Energy Gamma-ray Observatory eXplorer (AMEGO-X)}

\begin{figure}
  \begin{center}
    \includegraphics[width=0.62\linewidth]{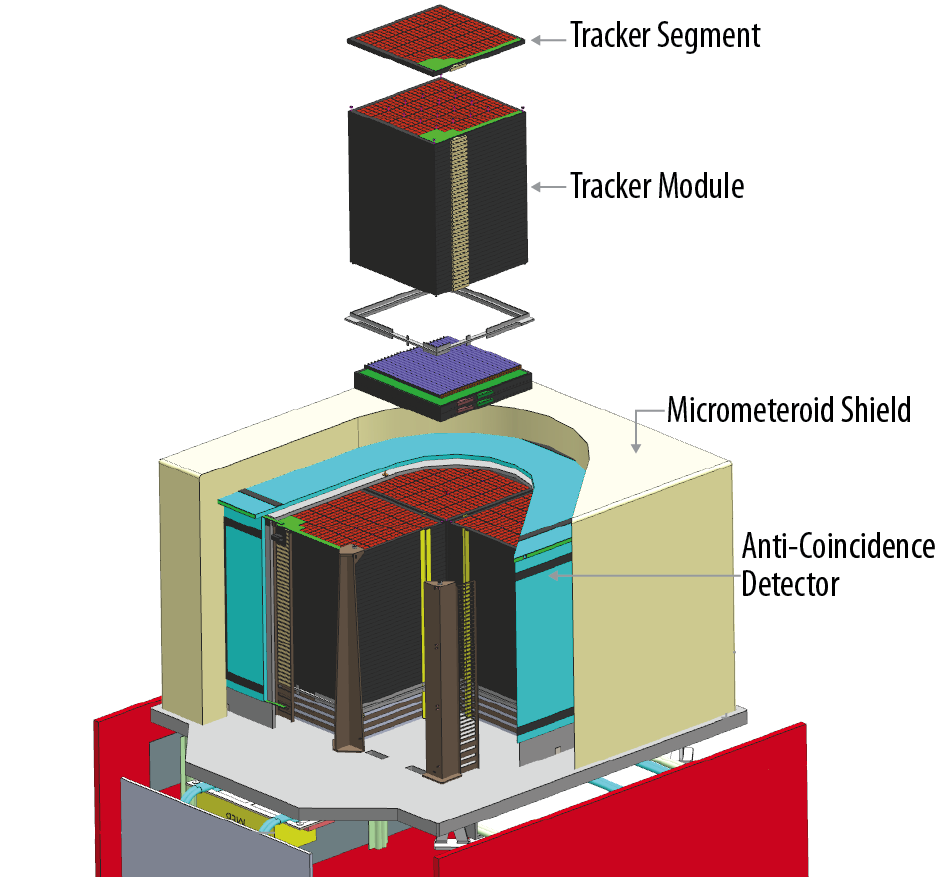}
    \end{center}
\caption{Exploded view of AMEGO-X.\label{AMEGO-X}}
\end{figure}

AMEGO-X is a mission concept proposed in NASA's 2020 MIDEX Announcement of Opportunity~\cite{Fleischhack:2021mhc}. It is designed to identify and characterize $\upgamma$-rays from extreme explosions and accelerators, the particle jet composition of actively accreting supermassive black holes~\cite{Lewis:2021roc}, the production and structure of jets from neutron star mergers~\cite{Martinez-Castellanos:2021bbl}, the diversity of particles accelerated by remnants of Galactic supernovae, and a host of other astrophysical events that produce all cosmic messengers~\cite{Negro:2021urm}. 
AMEGO-X will observe nearly the entire sky every two orbits during its three-year baseline mission, building up a sensitive all-sky map of $\upgamma$-ray sources and emissions. It will also access $>$50\% ($<$10\,MeV) and $>$20\% ($>$10\,MeV) of the sky instantaneously, maximizing transient detections and rapid alerts, openly distributed to the astrophysics communities. 

AMEGO-X will probe the medium energy $\upgamma$-ray band (25\,keV to over 1\,GeV) using a single instrument with sensitivity an order of magnitude greater than previous missions in this energy range that can be only realized in space. 
The $\upgamma$-ray Telescope, the sole instrument onboard, is comprised of three subsystems (Fig.~\ref{AMEGO-X}): a silicon pixel tracker, a Cesium Iodide (CsI) calorimeter, and a plastic anticoincidence detector.

If selected, AMEGO-X is planned to launch in 2028. AMEGO-X will observe nearly the entire sky every two orbits during its three-year baseline mission, building up a sensitive all-sky map of $\upgamma$-ray sources and emissions. It will also access $>$50\% ($<$10\,MeV) and $>$20\% ($>$10\,MeV) of the sky instantaneously, maximizing transient detections and rapid alerts, openly distributed to the astrophysics communities. 

\subsection{X-ray Experiments}

\subsubsection{Current Status}

Astrophysical observations performed with the current generation of X-ray instruments provide the leading direct experimental constraints on sterile neutrino decay over the complete keV-mass range, where sterile neutrinos could constitute DM~\cite{PhysRevLett.72.17},  and leading limits on ALPs for masses $<10^{-11}$\,eV.

Together with constraints from production in the early universe~\cite{Venumadhav:2015pla, Laine:2008pg} and structure formation~\cite{Cherry:2017dwu}, current X-ray constraints leave only a narrow window for sterile neutrinos to exist in the simplest models.
For masses  $<10$\,keV, the strongest limits come from focusing telescopes such as XMM-Newton and Suzaku~\cite{Ruchayskiy:2015onc, Malyshev:2014xqa, Tamura:2014mta} due to their large effective areas; however, these sensitivities decrease rapidly toward higher masses.
NuSTAR provides leading limits for masses \til$10-50$\,keV~\cite{Perez:2016tcq,Ng:2019a,Roach:2019ctw}. This sensitivity exploits the unique NuSTAR geometry, in which there is a large aperture for photons that bypass the focusing optics and, instead, arrive directly onto the detectors from off-axis angles \til$1-3^\circ$.  
Non-focusing instruments such as Fermi/GBM and INTEGRAL/SPI are sensitive to higher-energy photons and provide leading limits for masses $>50$\,keV~\cite{Ng:2015gfa, Boyarsky:2007ge}. 

In 2014, two groups~\cite{Bulbul:2014sua,Boyarsky:2014jta} reported the high-significance detection of a previously unknown X-ray line with rest-frame energy \til3.5\,keV (hereafter ``the 3.5-keV line'') in the XMM-Newton spectra of several galaxies and galaxy clusters. 
If interpreted as arising from the decay of a \til7-keV sterile neutrino dark-matter particle, the inferred mixing angle would lie in the range $\sin^2(2\theta) \approx (0.2-2)\times 10^{-10}$, depending on the choice of DM density profile. 
Nearly all operating X-ray observatories -- including XMM-Newton~\cite{Anderson:2014tza,Malyshev:2014xqa,Jeltema:2015mee,Ruchayskiy:2015onc,Gewering-Peine:2016yoj,Boyarsky:2018ktr,Dessert:2018qih,Dessert:2020hro,Bhargava:2020a,Foster:2021ngm}, Chandra~\cite{Horiuchi:2013noa,Hofmann:2016urz,Riemer-Sorensen:2014yda,Cappelluti:2017ywp,Hofmann:2019ihc,Sicilian:2020a}, Suzaku~\cite{Urban:2014yda,Tamura:2014mta,Sekiya:2015jsa,Bulbul:2016yop,Franse:2016dln}, NuSTAR~\cite{Neronov:2016wdd,Perez:2016tcq}, HaloSat~\cite{Silich:2021sra}, XQC~\cite{Figueroa-Feliciano:2015gwa} and Hitomi~\cite{Aharonian:2016gzq,Tamura:2018scp} -- have searched for this line in a variety of astrophysical targets. 
The 3.5-keV line has been detected with comparable interaction strength by
Chandra deep-field observations~\cite{Cappelluti:2017ywp}, 
XMM-Newton spectra of the Galactic halo~\cite{Boyarsky:2018ktr}, and
NuSTAR spectra of extragalactic survey fields~\cite{Neronov:2016wdd} (though the NuSTAR detection is consistent with a known instrument background feature ~\cite{Perez:2016tcq, Wik:2014}).
Alternative proposals for the origin of the 3.5-keV feature include plasma emission lines~\cite{Jeltema:2014qfa,Gu:2015gqm,Gu:2017pjy,Shah:2016efh}, novel plasma charge-exchange processes~\cite{Gu:2015gqm,Gu:2017pjy,Shah:2016efh}, and alternative dark-matter models~\cite{Cline:2014vsa,Conlon:2014xsa,Finkbeiner:2014sja,Cicoli:2014bfa,Brdar:2017wgy}.  
Recent high-statistics analyses of Chandra and XMM-Newton data~\cite{Sicilian:2020a,Dessert:2018qih,Foster:2021ngm} severely constrain the central value of the line strength, and are limited mainly by instrumental backgrounds and Galactic halo uncertainties~\cite{2020arXiv200406601B,2020arXiv200406170A}.  
To fully disentangle the 3.5-keV line from background features, input from future observatories will be critical.

Several recent analyses have used X-ray observations to probe low-mass ($<10^{-11}$\,eV) ALPs. 
The most stringent limits to date on low-mass ALPs come from the search in Chandra observations for X-ray spectral distortions in the active galactic nucleus NGC 1275 at the center of the Perseus cluster, induced by photon-axion conversion in the intra-cluster medium.  
These exclude values of the axion-photon coupling ${g_{a\gamma}>(6-8)\times10^{-13}\mathrm{\,GeV}^{-1}}$ (99.7\% C.L.) for masses $m_{a}<10^{-12}\mathrm{\,eV}$~\cite{chandra:2019uqt}. 
A similar analysis exploited Chandra's observations of the core of M87~\cite{m87:2017yvc}. 
However, these results could be significantly weakened depending on the relative magnitude of the regular and turbulent intra-cluster magnetic fields assumed~\cite{Libanov:2019fzq}.
Two recent analyses exploited NuSTAR hard X-ray observations of Betelgeuse~\cite{2021PhRvL.126c1101X} and of the  Quintuplet and Westerlund 1 super star clusters~\cite{Dessert:2020hro} to probe axions produced in stellar cores that are converted back into photons in the Galactic magnetic field. 

\paragraph{Chandra}
The Chandra X-ray observatory~\cite{2000SPIE.4012....2W} was launched in 1999. 
Chandra uses four pairs of nested mirrors, which are highly-polished integral shells, to deliver the fine angular resolution of 0.5\,arcsec. 
The Advanced CCD Imaging Spectrometer (ACIS) focal-plane instrument consists of ten CCD chips and provides images and spectral measurements in the energy range 0.2--10\,keV.
The High Resolution Camera (HRC) focal-plane instrument consists of two microchannel plates and provides images in the energy range of 0.1--10\,keV. 
In addition, the High Energy Transmission Grating Spectrometer (HETGS) and the Low Energy Transmission Grating Spectrometer (LETGS) provide fine energy resolution measurements for spectroscopy.

\paragraph{High Throughput X-ray Spectroscopy Mission X-ray Multi-Mirror Mission (XMM-Newton)}

The XMM-Newton observatory was launched in 1999~\cite{2001A&A...365L...1J}. The primary instruments are the three European Photon Imaging Cameras (EPIC), two MOS-charge-coupled-device (CCD) cameras, and a single pn-CCD camera. 
These cover an energy range of 0.15--15\,keV and a field-of-view of 30\,arcmin. 
The focusing optic consists of 58 nested Wolter Type-1 mirrors, manufactured using electroformed Ni replication and ranging in diameter from 306 to 700\,mm. 

\paragraph{IntErnational Gamma-Ray Astrophysics Laboratory (INTEGRAL)}

INTEGRAL~\cite{KUULKERS2021101629}, launched in 2002, consists of three science instruments. 
The Spectrometer of INTEGRAL (SPI), which provides imaging and spectroscopy from 20\,keV up to 8\,MeV, is a coded mask of hexagonal tungsten tiles above a detector plane made out of 19 germanium crystals.
The Imager onboard the INTEGRAL Satellite (IBIS) observes from 15\,keV to 10\,MeV with an angular resolution of 12\,arcmin. It consists of a 95$\times$95 mask of rectangular tungsten tiles above a forward plane of 128$\times$128 Cadmium-Telluride tiles, backed by a 64$\times$64 plane of Caesium-Iodide tiles. 
Dual JEM-X units operate from 3 to 35\,keV, using microstrip gas scintillators below a mask of hexagonal tiles to provide more precise imaging. 

\paragraph{Suzaku}

Suzaku~\cite{2007PASJ...59S...1M} operated from 2005 until 2015. 
It initially consisted of three co-aligned instruments.
X-ray Telescopes (XRTs), nested conical foil mirror assemblies, delivered spatial resolution of about 1.8' half-power diameter (HPD) over the band \til0.2--12\,keV.
An X-ray Imaging Spectrometer (XIS), consisting of X-ray sensitive imaging CCD cameras, was located in the focal plane of the XRTs. 
The second instrument was a non-imaging, collimated Hard X-ray Detector (HXD). The HXD used Gadolinium Silicate crystal (GSO) with Bismuth Germanate crystal (BGO) anti-coincidence counters to deliver low-background measurements over the wide energy band 10--700\,keV. 
The last instrument, the X-ray Spectrometer (XRS), was a mercury telluride (HgTe) microcalorimeter designed to deliver 6--7\,eV (FWHM) over the 0.3--12\,keV energy band; XRS did not operate due to a malfunction shortly after launch.

\paragraph{The Nuclear Spectroscopic Telescope Array (NuSTAR)}

NuSTAR launched in 2012, consists of two co-aligned telescopes and focal plane modules~\cite{Harrison:2013, Madsen:2015}. 
The telescopes are conic-approximation Wolter optics, each with 133 shells of radii 54--191\,mm. The depth-graded multi-layer coatings provide reflectivity up to 79\,keV, far beyond the \til10--12\,keV cutoff of previous focusing optics.
 Photons are focused onto solid state cadmium zinc telluride (CdZnTe) pixel detectors, providing a field-of-view approximately 13$\times$13\,arcmin$^2$ and an energy resolution of \til$0.4$\,keV for energies $<20$\,keV. 

\paragraph{Hitomi} 

Hitomi, launched in 2016, carried several X-ray detectors and a $\upgamma$-ray instrument onboard~\citep{Takahashi2018}. The non-dispersive microcalorimeter with 7\,eV resolution had a field-of-view of 3$\times$3\,arcmin$^{2}$ and was sensitive to 0.2--10\,keV band with an effective area of 210~cm$^{2}$ at 6\,keV~\citep{Kilbourne2016}. A new generation CCD detector, sensitive to 0.4--10\,keV X-ray bandpass with a large field-of-view (38$\times$38\,arcmin${^2}$), had an effective area of 360 cm$^{2}$ at 6\,keV and a moderate spectral resolution (200 eV at 6\,keV)~\citep{Tanaka2018}. Additionally, Hitomi carried a hard X-ray imager for sensitive imaging spectroscopy in the 5-80\,keV band~\citep{Nakazawa2018} and a non-imaging soft $\upgamma$-ray detector extending Hitomi’s energy band to 600\,keV~\citep{Tajima2018}. After an anomaly with the satellite operations, the Hitomi mission was lost after about 1.5\,months. Due to its short lifetime and prominent detection features in its early data, the mission ended before it could place meaningful constraints on the origin of the 3.5-keV line~\citep{Aharonian:2016gzq}.

\subsubsection{Near-term Future}

\paragraph{Extended ROentgen Survey with an Imaging Telescope Array (eROSITA)} 

\begin{figure}
\begin{center}
\begin{minipage}[t]{0.525\linewidth}
\includegraphics[width=1\linewidth]{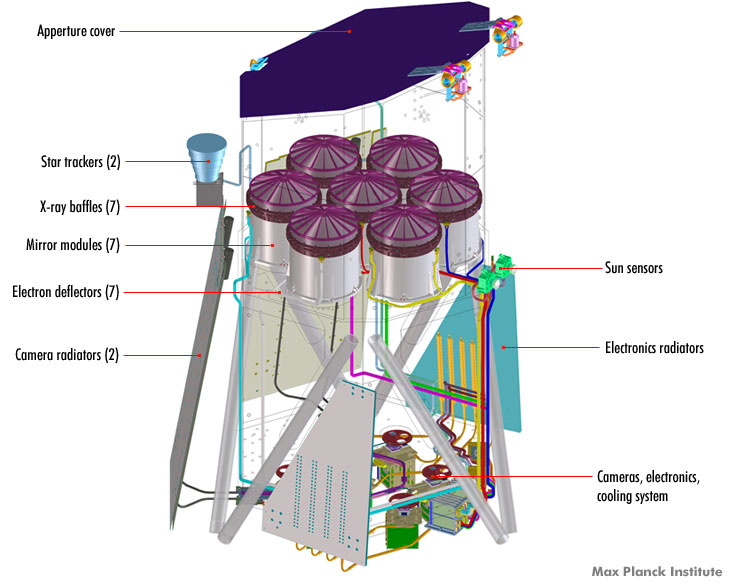}
\caption{Components of the eROSITA X-ray telescope on board Spectrum-Roentgen-Gamma Mission. \label{eROSITA}}
\end{minipage}
\hspace{0.03\linewidth}
\begin{minipage}[t]{0.415\linewidth}
    \includegraphics[width=1\linewidth]{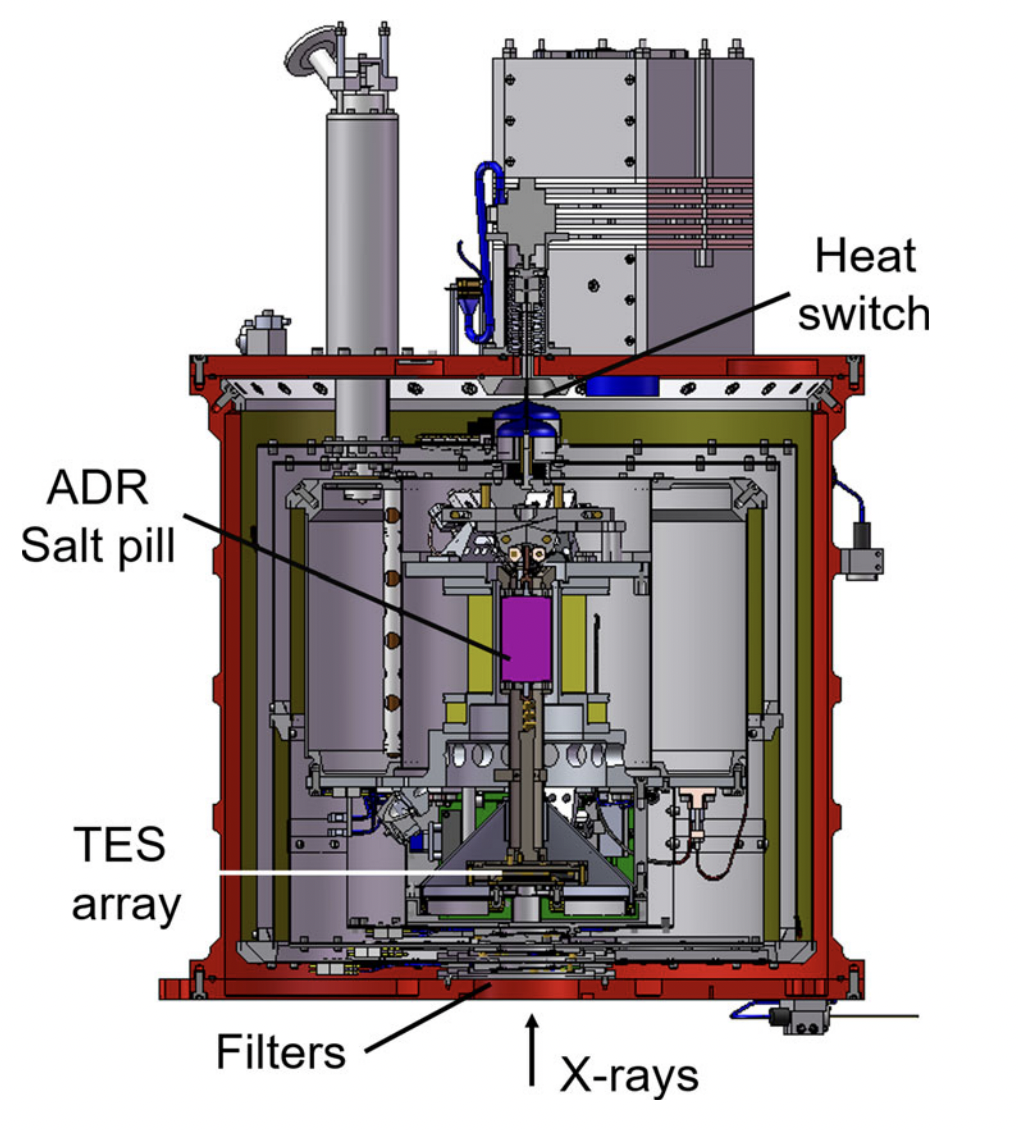}
\caption{Components of the Micro-X instrument \label{microX}}
\end{minipage}
\end{center}
\end{figure}

Launched in 2019, eROSITA, the primary soft X-ray instrument onboard the Russian-German Spectrum-Roentgen-Gamma (SRG) mission, scans the entire sky every six months~\cite{Sunyaev2021}. eROSITA (Fig.~\ref{eROSITA}) consists of seven co-aligned independent Walter-I type mirror modules and CCD detectors. The planned eight full-sky surveys are expected to be finalized in 2024. eROSITA carries seven independent X-ray telescope at a moderate spectral resolution (70--95\,eV at 1.49\,keV) in the 0.2--10\,keV band pass~\cite{Predehl2021}. The sensitivity of the survey is designed to be 20--30 times larger than the previous ROSAT All-Sky survey~\cite{Merloni2012}. The All-Sky Survey observations with eROSITA and complementary optical surveys will considerably extend the existing bounds in the next decade. The superb soft sensitivity (effective area of 367.9\,cm$^{2}$ at 1.49\,keV), wide-area coverage, and moderate spectral resolution will make the survey uniquely suited for searches for weak X-ray lines and spectral distortions due to ultralight axions photon conversion both in the Milky Way observations and through stacking method of the X-ray emission from other dark matter-dominated objects~\cite{Dekker2021, Bulbul2021, Barinov2021, Ando2021}.

\paragraph{X-ray Imaging and Spectroscopy Mission (XRISM)} 

XRISM is scheduled to launch in 2023 and will carry a non-dispersive, high-resolution X-ray microcalorimeter with 7\,eV energy resolution sensitive in the soft X-ray bandpass between 0.3--12\,keV~\cite{Tashiro2020}. Due to its small field of view (2.9$^{\prime}\times$2.9$^{\prime}$), moderate effective area (\til160\,cm$^{2}$ at 1\,keV), and large angular resolution (1.7\,arcmin (HPD), XRISM will require long exposure times in the order of several million seconds of dark matter dominated objects to detect weak lines that may result from heavy neutron decays or axion conversion signal. Combining all cluster observations with XRISM over several years will offer a chance to detect a possible signal from dark matter if the energy calibration is accurate in the relevant X-ray band~\cite{XRISM2020, Zhong2020}.
    
\paragraph{Micro-X} 

Micro-X~\cite{2020JLTP..199.1072A} is an X-ray spectroscopy sounding rocket mission with a target bandpass of 0.5--10\,keV, and target energy resolution of 3\,eV. Micro-X consists of a 128-pixel Transition Edge Sensor (TES) array with SQUID readout (Fig.~\ref{microX}) . For its first flight in 2018~\cite{2020JLTP..199.1062A}, the instrument was in imaging configuration. Although a rocket pointing error prevented taking images of the target, the supernova remnant Cassiopeia A, valuable engineering data was collected. The payload is proposed to be modified to widen the field-of-view and optimize the higher-energy response~\cite{2020JLTP..199.1072A}. With this modification, Micro-X would have competitive sensitivity to sterile neutrino dark matter even for the short 300\,s flight-times available on a rocket. In addition, the fine energy resolution would allow Micro-X to perform velocity spectroscopy of any detected line, helping distinguish between dark matter signals and atomic backgrounds. 

\subsubsection{Proposed Future Missions}
\paragraph{Advanced Telescope for High ENergy Astrophysics (Athena)} 

\begin{figure}
  \begin{center}
    \includegraphics[width=1\linewidth]{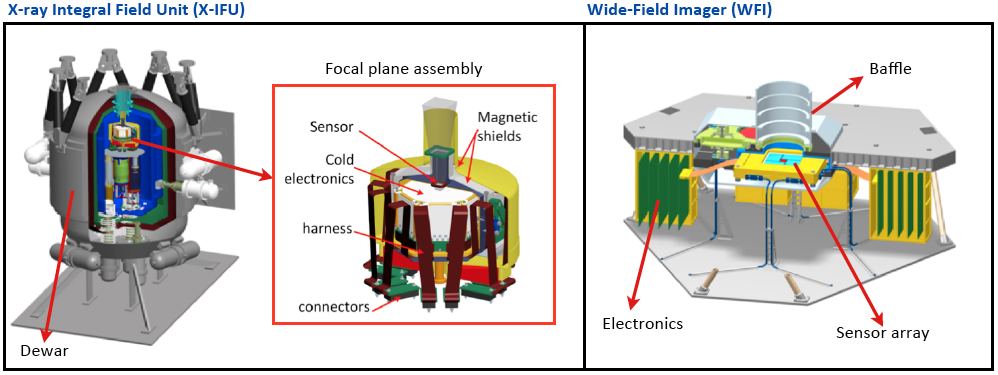}
    \end{center}
    
\caption{The two X-ray instruments XIFU and WFI on board Athena. \label{athena}}
\end{figure}

The next-generation X-ray observatory Athena, selected by ESA, is scheduled to launch in the early 2030s~\cite{Nandra2013}. Athena (Fig.~\ref{athena}) will be carrying two X-ray instruments: Wide Field Imager (WFI) \citep{Meidinger2017} and X-ray Integral Field Unit (XIFU) \citep{Barret2018}. The WFI instrument is composed of DEPFET detectors sensitive to 0.2-10\,keV band with a large field-of-view $40^{\prime} \times 40^{\prime}$ and has a moderate spectral resolution of $<80$\,eV at 1\,keV, while XIFU provides high-resolution spectroscopy in the 0.2--12\,keV energy band (2.5\,eV FWHM energy resolution out to 7\,keV). The requirement of the spatial resolution of the mirror is 5\,arcsec with an effective area of 1~m$^{2}$ (at 1\,keV) is 45 times larger than the one of XRISM/Resolve~\cite{Barret2018}. The Athena-WFI, owing to its large field of view and surveying capability, will be able to scan the Milky Way Halo out to large angular scales in achievable short exposures while resolving the smaller substructures that might contribute astrophysical and detector background with its fine angular resolution. The Athena-XIFU calorimeter will resolve complex line structures with its large weak and narrow line sensitivity and effective area, i.e., at least an order of magnitude better at 1\,keV than the XRISM/Resolve instrument. Athena-XIFU will have the capability to detect weak dark matter decay lines and spectral distortions due to axion-photon conversion in nearby clusters of galaxies while measuring the velocity dispersion of dark matter particles~\cite{Bulbul:2014sua, Zhong2020}. 

\section{Cosmic-ray Probes}\label{sec:cosmicrays}
\label{sec:CR}

\textit{Contributors (alphabetical order): Mirko Boezio, Philip von Doetinchem}
\bigskip

This section summarizes the current status and outlook of cosmic-ray probes with sensitivity to dark matter. As discussed in Sec.~\ref{s-intro}, the focus for the dark matter search with charged particles is on measuring antiparticle spectra, namely positrons and light antinuclei (antiprotons, antideuterons, antihelium nuclei), because of their very low flux from standard astrophysical sources. 
However, most of the following experiments also have substantial sensitivity to particle fluxes. A precise understanding of particle fluxes is vital for constraining cosmic-ray propagation parameters, and, as such, they directly impact the interpretation of the antiparticle fluxes. 
Therefore, the capabilities for particle fluxes are pointed out where applicable.

\subsection{Current Status}

\paragraph{Balloon-borne Experiment with Superconducting Spectrometer (BESS)\label{s-bess}}

\begin{figure}
  \begin{center}
    \includegraphics[width=1\linewidth]{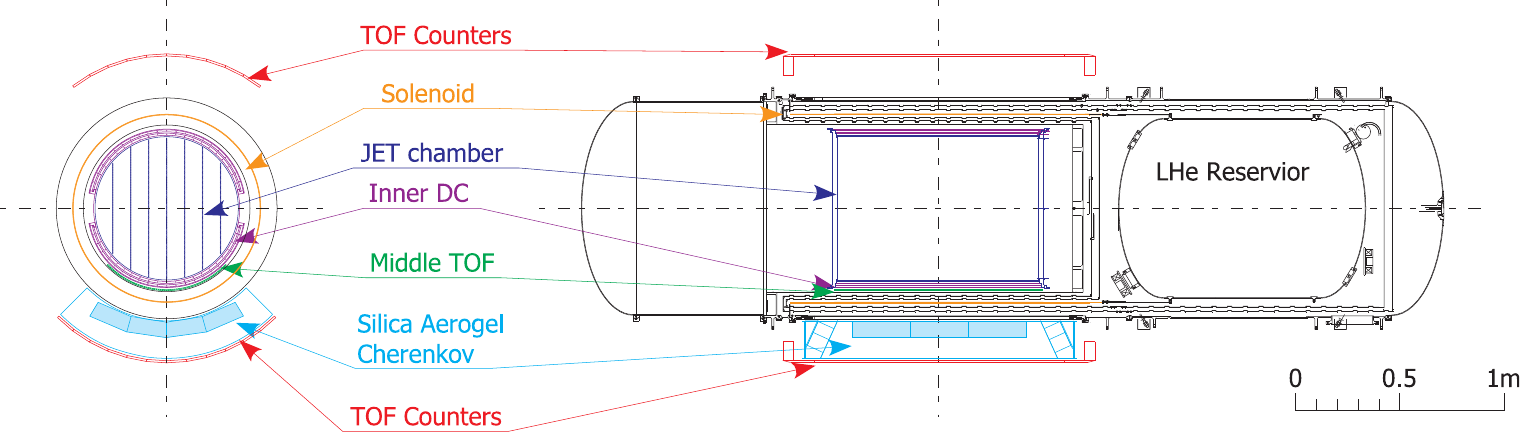}
    \end{center}
\caption{\label{f-fig1} The BESS-Polar II experiment.}
\end{figure}

The BESS program (1993-2008)~\cite{2013AdSpR..51..227B}, culminating in the BESS-Polar instrument, exploits particle tracking in a solenoidal magnetic field to identify antimatter. 
The original BESS-Polar experiment flew over Antarctica in late 2004. The BESS-Polar II experiment collected 24.5\,days of Antarctic flight data from December 2007 to January 2008~\cite{besspbar,2012PhRvL.108m1301A}. BESS-Polar II, shown in Fig.~\ref{f-fig1},  consists of a 0.8\,T solenoidal magnet, filled by inner drift chambers (IDC) and a jet-type drift tracking chamber (JET), and surrounded by an aerogel Cherenkov counter (ACC) and a time-of-flight (TOF) system composed of scintillation counter hodoscopes. These components are arranged in a coaxial cylindrical geometry, providing a sizeable geometric acceptance of 0.23\,m$^{2}$\,sr. The threshold rigidities for antiproton and antideuteron are 3.8\,GV and 7.6\,GV, respectively. In addition, a thin scintillator middle-TOF with a timing resolution of 320\,ps is installed between the central tracker and the solenoid to detect low-energy particles that cannot penetrate the magnet wall. BESS-Polar II provided an antiproton spectrum in the energy range from approximately 200\,MeV/$n$ to 3\,GeV/$n$~\cite{besspbar}; 
below 500\,MeV/$n$, this is the highest-precision antiproton measurement currently available. 
It also set an exclusion limit for antihelium of $1.0\times10^{-7} (\text{m}^2\text{s sr\,GeV}/n)^{-1}$ in the range of 1.6--14\,GV~\cite{2012PhRvL.108m1301A}, {based on the specific assumption that antihelium and helium have the same spectral shape. It needs to be noted that this assumption is generally not correct if the production mechanisms for helium and antihelium are different.}
The last published antideuteron results relied on the previous BESS flights, which took place between 1997 and 2000, setting an exclusion limit at 95\% confidence level of $1.9\times10^{-4}(\text{m}^2\text{s sr\,GeV}/n)^{-1}$ in the range of 0.17--1.15\,GeV/$n$~\cite{Fuke:2005it}. 
Extended BESS-Polar II antiproton and the antideuteron analyses are currently ongoing, using the middle-TOF that lowers the energy range to about 100\,MeV/$n$.

\paragraph{Payload for Antimatter Matter Exploration and Light-nuclei Astrophysics (PAMELA)}

\begin{figure}
  \begin{center}
    \includegraphics[width=0.5\linewidth]{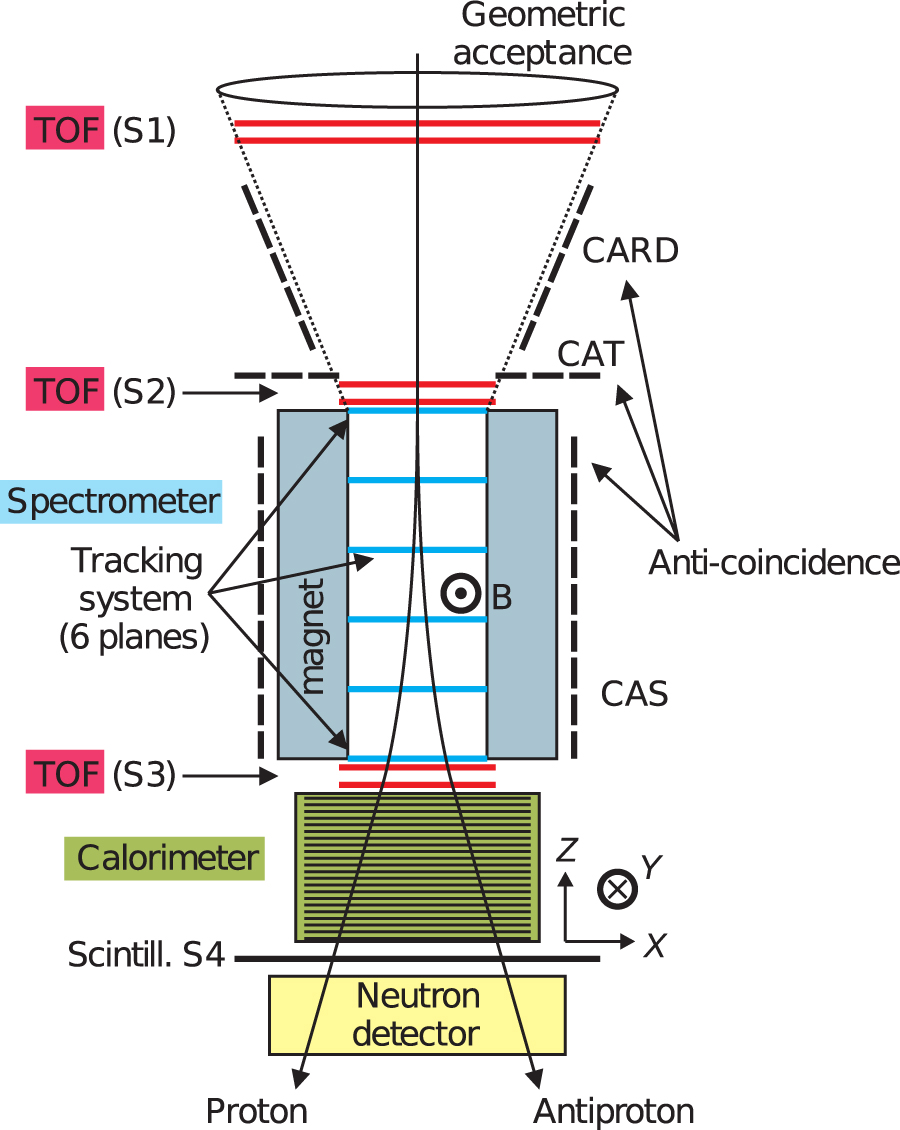}
    \end{center}
\caption{\label{f-pamela_schema} Layout of PAMELA.}
\end{figure}

The space-borne experiment PAMELA (Fig.~\ref{f-pamela_schema}) operated between 2006 and 2016. The PAMELA apparatus was designed to study charged particles in cosmic radiation, with a particular focus on antiparticles for searching antimatter and signals of dark matter annihilation~\cite{pamelaexp}. The apparatus comprised a time-of-flight system, a silicon-microstrip magnetic spectrometer, a silicon-tungsten electromagnetic calorimeter, an anticoincidence system, a shower tail catcher scintillator, and a neutron detector~\cite{Adriani_PhysRep_2014,PAMELA:2017bna}. The experiment conducted long-duration measurements of the cosmic radiation over an extended energy range from tens of\,MeV up to \til1\,TeV, including important new results on the antiparticle component of cosmic radiation~\cite{pamela, Adriani:2010ib, 2013PhRvL.111h1102A, Adriani_PhysRep_2014}. For instance, PAMELA detected a rising positron fraction, and the extension of positron and electron measurements to lower energies provided clear evidence of a charge-sign dependence in the propagation of charged particles in the heliosphere~\cite{Adriani:2016uhu,PAMELA:2017bna}. 
Furthermore, PAMELA significantly improved the existing experimental antiproton data, extending both the energy range and the available statistics in the range from 60\,MeV to 350\,GeV~\cite{Adriani:2008zq,2010PhRvL.105l1101A,Adriani:2012paa,Adriani_PhysRep_2014}.
Finally, the search for antihelium nuclei resulted in a stringent upper limit for their existence~\cite{Mayorov:2011zz, Adriani_PhysRep_2014, PAMELA:2017bna}.

\paragraph{Alpha Magnetic Spectrometer (AMS-02)\label{s-ams}}

\begin{figure}
  \begin{center}
    \includegraphics[width=0.55\linewidth]{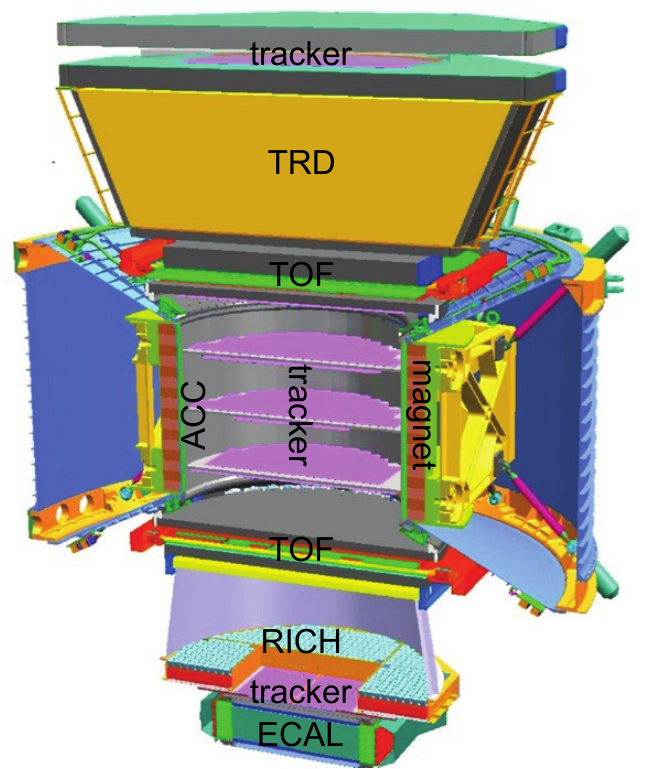}
    \end{center}
\caption{\label{f-fig2} Layout of AMS-02.}
\end{figure}

AMS-02 is a multi-purpose cosmic-ray detector operating on the International Space Station (ISS) since May 2011. Thus far, it has recorded about 200\,billion triggered events~\cite{ams4}. AMS-02 is planned to operate until the end of the lifetime of the ISS. In contrast to the high-statistics spectral measurements of other cosmic-ray species, the antideuteron and antihelium studies of AMS-02 are focused on a first-time discovery. AMS-02 follows the principle of typical magnetic spectrometer particle physics detectors, with particle identification that relies on combining signals from an array of sub-detectors, as shown in Fig.~\ref{f-fig2}. For antinuclei analyses, the transition radiation detector (TRD) suppresses low-mass particles such as electrons, pions, and kaons. The time-of-flight (TOF) system provides the main trigger and determines the particle's velocity up to $\beta\approx0.8$. The particle momentum can be extracted from its trajectory in the approximately 0.15\,T solenoidal magnetic field. 
In the high-velocity region, two different types of ring imaging Cherenkov (RICH) counters are used (NaF and aerogel) for the velocity measurement. 

A summary of the AMS-02 particle spectra that were measured so far can be found here ~\cite{AGUILAR20211}. Since the publication of ~\cite{AGUILAR20211}, results on Sodium, Aluminum, Nitrogen, Fluorine, and Iron were also added~\cite{PhysRevLett.126.041104, PhysRevLett.126.081102,PhysRevLett.127.021101, PhysRevLett.127.271102}. The particle spectra are critical for understanding the transport of cosmic rays in our Galaxy.

Furthermore, AMS-02 conducts precision positron measurements and observed a significant excess in the positron spectrum at about 25\,GeV and a sharp drop-off at about 284\,GeV. These properties are not explained by traditional cosmic-ray models~\cite{PhysRevLett.122.041102}. Concerning antinuclei, the AMS-02 collaboration has published the most precise antiproton spectrum in the range 1--525\,GV, based on $5.6\times10^5$ antiproton events~\cite{PhysRevLett.117.091103, AGUILAR20211}. Moreover, AMS-02 showed several candidate events with mass and charge consistent with antihelium at conferences~\cite{antihe,antihe2}. The analyses of antinuclei, including antideuterons, are ongoing while more data are collected.

After the upgrade of the AMS-02 tracker cooling pumps, it is planned that AMS-02 will continue taking data until the end of the lifetime of the International Space Station. Furthermore, an upgrade of the silicon tracker is currently under construction and will be installed by 2024, leading to a significant acceptance increase. Continuing the measurements will extend the energy range of positrons towards higher energies, improve the accuracy of the spectrum, and increase the sensitivity for searches for antimatter.

\paragraph{Calorimetric Electron Telescope (CALET)}

\begin{figure}
\begin{center}
\begin{minipage}[t]{0.46\linewidth}
    \includegraphics[width=1\linewidth]{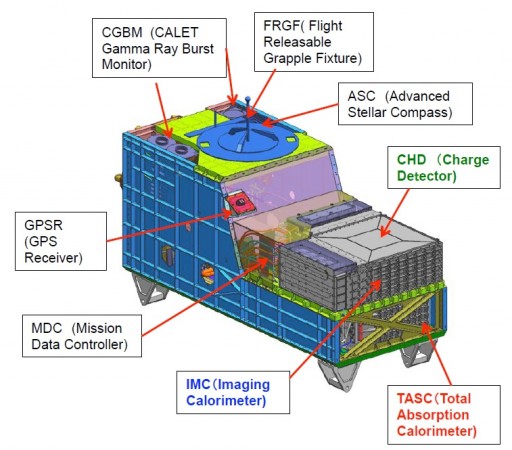}
\caption{\label{f-calet} Layout of the CALET experiment.}
\end{minipage}
\hspace{0.05\linewidth}
\begin{minipage}[t]{0.46\linewidth}
    \includegraphics[width=1\linewidth]{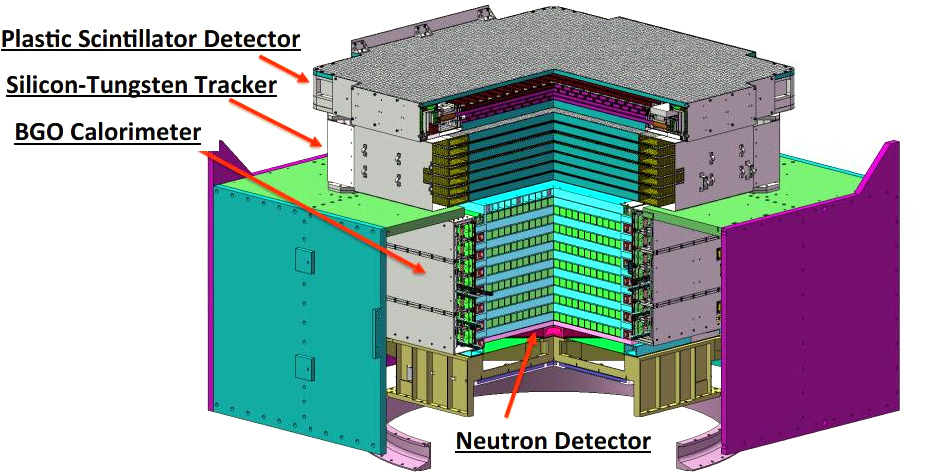}
\caption{\label{f-dampe} Layout of the DAMPE experiment.}
\label{f-Present}
\end{minipage}
\end{center}
\end{figure}

The CALET experiment (Fig.~\ref{f-calet}) has been installed on the International Space Station since 2015. It is equipped with a calorimeter and $\upgamma$-ray burst monitor but not a magnet. CALET already obtained the electron-positron spectrum up to about 5\,TeV, the proton spectrum up to about 10\,TeV, and the carbon,  oxygen, and iron spectra up to about 2\,TeV~\cite{CALET:2019bmh,Adriani:2020wyg,CALET:2021fks}. CALET will continue to operate until at least the end of 2024.

\paragraph{Dark Matter Particle Explorer (DAMPE)}

The DAMPE experiment (Fig.~\ref{f-dampe}) has been operating in space since late 2015, intending to measure cosmic-ray spectra above the TeV range. It consists of a plastic scintillator detector, a silicon tracker, a calorimeter, and a neutron monitor but does not have a magnet. DAMPE measurements of the electron-positron spectrum indicate a break in the TeV region~\cite{DAMPE:2017fbg}. Furthermore, DAMPE places limits on dark matter annihilation into monochromatic $\upgamma$-rays~\cite{LIANG2021}. 

\subsection{Near-term Future}

\paragraph{General Antiparticle Spectrometer (GAPS)\label{s-gaps}}

\begin{figure}
  \begin{center}
    \includegraphics[width=0.82\linewidth]{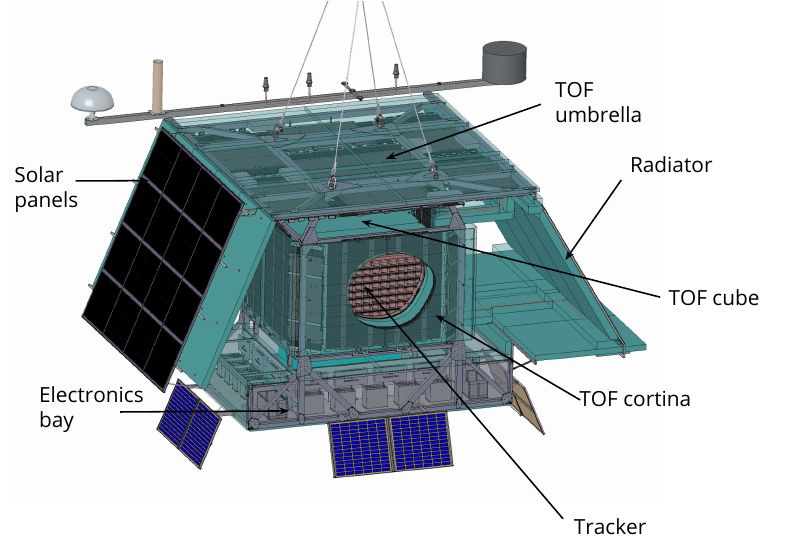}
    \end{center}
\caption{\label{f-fig3} Overview of the GAPS experiment.}
\end{figure}

The GAPS experiment is optimized for low-energy ($<0.25$\,GeV/$n$) cosmic-ray antinuclei~\cite{Aramaki2015}. The experiment, shown in Fig.~\ref{f-fig3}, consists of ten planes of semiconducting Si(Li) strip detectors surrounded by a plastic scintillator time-of-flight (TOF) system. GAPS is currently preparing for its first Antarctic long-duration balloon (LDB) flight, with the baseline sensitivity to antideuterons projected after a total of three Antarctic LDB flights.

GAPS relies on a novel particle identification technique based on exotic atom formation and decay~\cite{Aramaki2015}, in which antinuclei slow down and eventually annihilate within the detector.
The identification of antinuclei makes use of the simultaneous occurrence in a narrow time window of X-rays of characteristic energy and nuclear annihilation products, providing high rejection power to suppress non-antiparticle background and identify the antinucleus species. 
This exotic atom detector design yields a large grasp compared to typical magnetic spectrometers and allows for identifying antiproton, antideuteron, and antihelium cosmic rays. GAPS will provide a precision antiproton spectrum for the first time in the low-energy range below $0.25$\,GeV/$n$ and provide sensitivity to antideuterons that is about two orders of magnitude better than the current BESS limits. Though the instrument is optimized for antideuterons, the exotic atom detection technique is also sensitive to antihelium nuclei signatures~\cite{saffoldanti}. Due to the higher charge, the antihelium analysis is even less affected by antiproton backgrounds than the antideuteron analysis, which allows for a competitive antihelium sensitivity in the low-velocity range. Beyond this initial flight program, the GAPS collaboration is already developing a vision for an upgraded payload, e.g., suitable for the NASA Pioneers program, to further increase sensitivity by about a factor of 5, compared to three LDB flights.

\paragraph{High Energy Light Isotope Experiment (HELIX)}

\begin{figure}
  \begin{center}
    \includegraphics[width=0.4\linewidth]{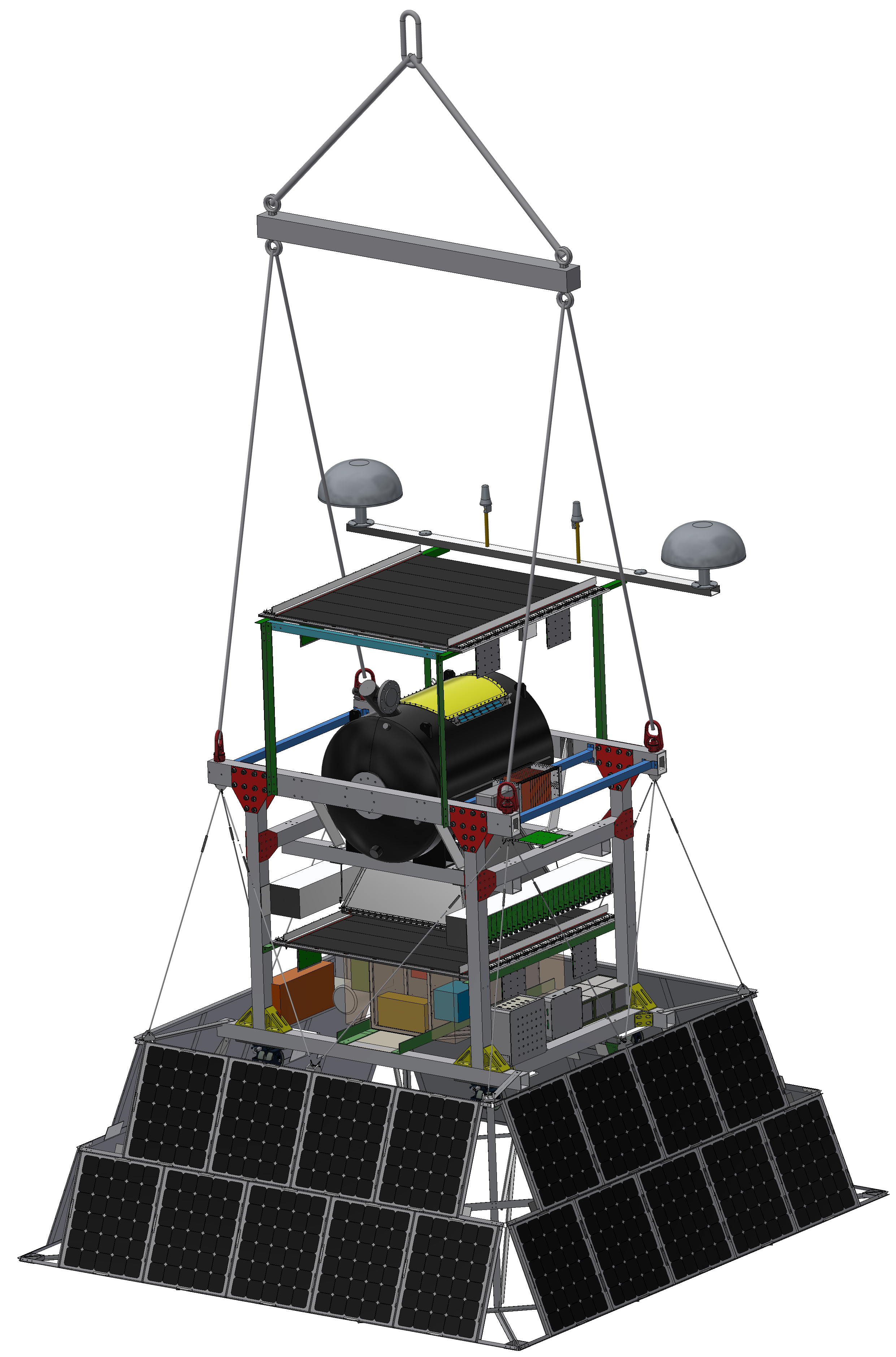}
    \end{center}
\caption{\label{f-helix} Overview of the HELIX experiment.}
\end{figure}

The HELIX instrument (Fig.~\ref{f-helix}) is a magnet spectrometer system consisting of a 1\,T superconducting magnet, a drift chamber tracker, and an aerogel RICH ~\cite{f2f8e111be344c99b12978ac18b696ce}. HELIX is designed to fly as a high-altitude long-duration balloon payload and is preparing for its first flight. The main focus of HELIX is the measurement of the Beryllium-10-to-Beryllium-9 ratio in the energy range of about 1--10\,GeV/$n$, which is an essential ingredient to constrain cosmic-ray propagation models by determining the diffusion time of cosmic rays in our Galaxy. The target for the first launch is spring 2023.

\subsection{Proposed Future Missions}

\paragraph{Gamma-ray and AntiMatter Survey (GRAMS)}
\label{sec:GRAMS}

The GRAMS experiment (Fig.~\ref{fig:GRAMS}) is a novel instrument designed to simultaneously target both astrophysical $\upgamma$-rays with\,MeV energies (Sec.~\ref{s-mevvision}) and antimatter signatures of DM~\cite{Aramaki:2019bpi}. The GRAMS instrument consists of a liquid-argon time projection chamber (LArTPC) surrounded by plastic scintillators. The LArTPC is segmented into cells to localize the signal, an advanced approach to minimize coincident background events in the large-scale LArTPC detector. 

The GRAMS concept potentially allows for a larger instrument since argon is naturally abundant and low cost than current experiments that rely on semiconductor or scintillation detectors. GRAMS is proposed to begin as a balloon-based experiment as a step forward to a satellite mission.

GRAMS has been developed to become a next-generation search for antimatter signatures of DM. The detection concept is similar to GAPS's, relying on exotic atom capture and decay. However, as the LArTPC detector can provide an excellent 3-dimensional particle tracking capability with nearly no dead volume inside the detector, the detection efficiency can be significantly improved while reducing the ambiguity of antimatter measurements, which is crucial for the discovery of rare events. GRAMS could investigate dark matter models that could potentially explain the Fermi Galactic Center excess and the AMS-02 antiproton excess~\cite{Aramaki:2019bpi}. 

\paragraph{AntiDeuteron Helium Detector (ADHD)}

The AntiDeuteron Helium Detector (ADHD) aims to use the distinctive signature of delayed annihilation of antinuclei in helium to identify cosmic antimatter species. 
The typical lifetime for stopped antideuterons in matter is on the order of picoseconds, similar to that of stopped antiprotons. However, the existence of long-lived (on the order of microseconds) metastable states for stopped antiprotons in helium targets has been measured~\cite{PhysRevLett.67.1246}. These metastable states in helium have also been measured for other heavy negative particles, such as pions and kaons~\cite{1992PhRvA..45.6202N, 1989PhRvL..63.1590Y}. The theoretical description of this effect predicts that the lifetimes of these metastable states increase quadratically with the reduced mass of the system, i.e., a larger delay of the annihilation signature is expected for antideuteron capture in helium than for antiproton capture~\cite{osti_4029624, PhysRevLett.23.63, PhysRev.188.187, PhysRevA.6.2488, PhysRevA.51.2870}.

The inner portion of the ADHD layout contains a helium calorimeter (HeCal), consisting of a spherical thermoplastic vessel filled with scintillating helium. Helium gas is a fast UV scintillator, having a light yield similar to other fast plastic or liquid scintillators and capable of providing nanosecond timing performance~\cite{doi:10.1063/1.3665333}. The HeCal is surrounded by a time-of-flight system consisting of plastic scintillator bars, which provide velocity and charge measurements via ionization energy loss.
Combining information on the velocity and energy depositions measured by the TOF, the prompt and delayed energy measured by the HeCal, and the reconstructed event topology, it is possible to identify a single antideuteron over $10^3$ background antiprotons. 

\paragraph{Next-generation Superconducting Magnetic Spectrometers}

\begin{figure}
  \begin{center}
    \includegraphics[width=0.8\linewidth]{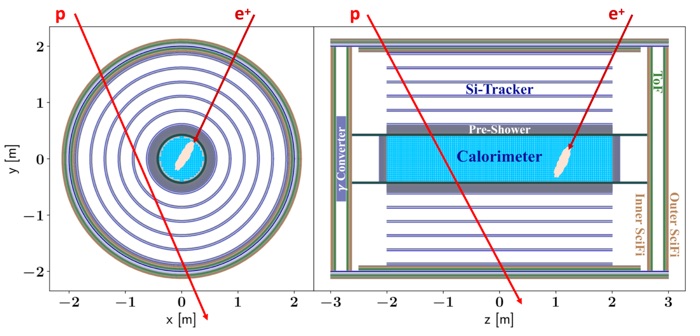}
    \end{center}
\caption{\label{f-ams-100} Concept of the AMS-100 experiment.}
\end{figure}

The conceptualized Antimatter Large Acceptance Detector In Orbit (ALADInO)~\cite{Battiston:2021org,instruments6020019} and A Magnetic Spectrometer (AMS-100)~\cite{2019NIMPA.94462561S} experiments are next-generation magnetic spectrometers with significantly increased acceptance as compared to AMS-02.

A possible ALADInO detector is equipped with a superconducting magnetic spectrometer, a 3-dimensional imaging calorimeter, and a time-of-flight with approximately 1\% resolution for efficient deuteron-proton separation. The detector concept maximizes the collection power using a toroidal magnetic configuration. As a result, ALADInO will reach about two orders of magnitude more separation power for particles-antiparticle separation than current experiments.

AMS-100 is an idea for a space-based international platform for precision particle astrophysics that will drastically improve on existing measurements of cosmic rays and $\upgamma$-rays~\cite{2019NIMPA.94462561S}. The critical component of the instrument (Fig.~\ref{f-ams-100}) is a thin, large-volume, high-temperature superconducting solenoid magnet. When instrumented with silicon-strip and scintillating fiber technologies, the spectrometer can achieve a maximum detectable rigidity of 100\,TV. In addition, a deep central calorimeter provides energy measurements and particle identification. The magnet and detector systems will be designed with no consumables, allowing for an extended 10-year payload lifetime at its thermally-favorable orbital location at Sun-Earth Lagrange Point 2. 

\section{Collider Experiments}\label{sec:colliders}

\textit{Contributors (alphabetical order): Philip von Doetinchem, Fiorenza Donato}
\bigskip

Understanding the standard astrophysical backgrounds is crucial to identifying a dark matter signal in photon or cosmic-ray data~\cite{SnowmassCF1WP6}. A key ingredient is the production cross sections for photons and antiparticles in interactions with the interstellar gas. For instance, a new antiproton production cross section parameterization~\cite{Winkler:2017xor}, considering PHENIX~\cite{Adare:2011vy}, STAR~\cite{Abelev:2006cs,Abelev:2008ab}, CMS~\cite{Khachatryan:2011tm,Chatrchyan:2012qb,Zsigmond:2012vc,H.VanHaevermaetfortheCMS:2016uir}, and ALICE~\cite{Aamodt:2011zza, Aamodt:2011zj} data, together with a new parameterization of the boron-to-carbon ratio, resolved the discrepancy between the prediction and the antiproton flux measured by AMS-02 at high energies. Furthermore, the modeling of $\upgamma$-ray emission in the Galaxy is based on the inverse Compton scattering of $e^\pm$ off the interstellar radiation field and the hadronic photon production in the nuclei cosmic ray collisions via $\pi^0$ decay. 
More data on the inclusive cross sections $p+p \rightarrow \pi^0 + X$ and $p+\text{He} \rightarrow \pi^0 + X$ would be desirable 
for the modeling of the Galactic emission~\cite{2012ApJ...750....3A, Kachelriess:2014mga}. 
The interpretation of the plethora of data taken by the Fermi-LAT regarding point sources and diffuse emission relies on cross sections still affected by considerable uncertainties 
~\cite{2011A&A...531A..37D,2015ApJ...806..240C,2011A&A...531A..37D}. Models for cross sections should not exceed 5\% uncertainty, both in normalization and shape. A strong need also exists for the measurement of the $ p + p \rightarrow  e^+ + X$ and $ p + \text{He} \rightarrow  e^+ + X$ inelastic cross sections, for positron energies from hundreds of\,MeV up to about 500\,GeV~\cite{2010A&A...524A..51D}, where an excess in the data is observed with respect to the secondary emission. The current models for secondary $e^+$ fluxes at Earth 
~\cite{1998ApJ...493..694M,2010A&A...524A..51D} rely on old data~\cite{Kelner:2006tc} or Monte Carlo simulations~\cite{Kamae:2006bf}. The uncertainty due to the cross sections is about a factor 2~\cite{2010A&A...524A..51D}.   The following reviews the current status and motivation for improving photon and cosmic ray production cross sections with ground-based experiments.

\subsection{Current Status}

The uncertainties on antiproton production from protons interacting with heavier nuclei are larger than those from $p$--$p$ interactions, because very few direct measurements exist. Instead, these cross sections are calculated by rescaling the $p$--$p$ cross sections. At lower energies, new $p$--$p$ data ($\sqrt{s}=7.7, 8.8, 12.3, 17.3$\,GeV) became available from NA61/SHINE in 2017~\cite{Aduszkiewicz:2017sei}. In addition, the first antiproton production cross section in $p$--He collision from LHCb at $\sqrt{s}=110$\,GeV was published~\cite{Aaij:2018svt}. 

For heavier antinuclei made of multiple antinucleons, it is essential to note that typically every production process should also produce antiprotons in much higher quantity, with each additional antinucleon reducing the yield by a factor of approximately 1000. Therefore, a correlation exists between the measured cosmic-ray antiproton flux and the predicted heavier antinuclei fluxes. However, the heavier antinuclei formation processes are not well constrained~\cite{Gomez-Coral:2018yuk} and allow for a wide span of predicted fluxes that are still compatible with the observed antiproton flux. It is an important question whether antinuclei are produced in collision via freeze-out from a quark-gluon plasma (the statistical thermal model) or at a later stage via coalescence of individual antinucleons (e.g.,~\cite{Mrowczynski:2016xqm}). 

Important constraints for the antinuclei flux from dark matter annihilations are coming from the values of the diffusion coefficient, its rigidity dependence, and the Galactic halo size, which directly scales the observable flux~\cite{Donato:2003xg}. Fits of cosmic-ray nuclei data for secondary-to-primary ratios (e.g., Li/C, Li/O, Be/C, Be/O, B/C, B/O) are important to constrain propagation models~\cite{1990acr..book.....B,2007ARNPS..57..285S,2017ApJ...840..115B,Reinert:2017aga,2018ApJ...854...94B,2018ApJ...858...61B,Boudaud:2019efq,2019arXiv191103108B}. However, this approach is hampered by uncertainties in the production  cross sections at the level of 10--20\%~\cite{0067-0049-144-1-153,2010A&A...516A..67M,2015A&A...580A...9G,Tomassetti:2017hbe,Genolini:2018ekk,Evoli:2019wwu}. These uncertainties propagate directly into predictions for the fluxes. 
These secondary-to-primary flux ratio fits are also somewhat degenerate in the ratio of the diffusion coefficient normalization to the halo size, which can be broken if radioactive secondaries like Beryllium-10 are used~\cite{2007ARNPS..57..285S}. 

\subsection{Near-term Future}

Antiproton cross section uncertainties in the energy range of AMS-02 are at the level of 10--20\%, with higher uncertainties for lower energies. For energies lower than the AMS-02 range, relevant for the GAPS experiment, a significant uncertainty on the source term from cross section normalization and shape exist. Future measurements at low center-of-mass energies ($<7$\,GeV) could improve these antiproton flux uncertainties~\cite{Donato:2017ywo}. 
Improving antiproton cross section measurements is also relevant for a precision understanding and antinuclei formation. Furthermore, it is necessary to use different collision systems with energies closer to the production threshold of light antinuclei to understand their production in the Galaxy.

Improvement in the accuracy of the production cross sections of secondary cosmic rays from MeV/$n$ up to at least TeV/$n$ energies is key to for the understanding of cosmic-ray antinuclei as well because they are critical for the determination of Galactic propagation properties to which all charged particles obey
~\cite{2020A&A...639A..74W, Korsmeier:2021bkw}. Furthermore, in the future, deuterons, helium-3, and sub-iron nuclei can be used to study propagation effects on different spatial scales~\cite{Trotta:2010mx}.

\paragraph{SPS Heavy Ion and Neutrino Experiment (NA61/SHINE)}

The fixed-target experiment NA61/SHINE at the Super Proton Synchrotron (SPS) at CERN is a hadron spectrometer capable of studying collisions of hadrons with different targets over a wide incident beam momentum~\cite{na61}. The detector consists of different subdetectors for the particle identification. NA61/SHINE already recorded $p$--$p$ interactions with beam momenta from 13 to 400\,GeV/$c$, and also collected data for other hadron interactions, including $p$--C, $\pi^+$--C, Ar--Sc, $p$--Pb, Be--Be, Xe--La, Pb--Pb at different energies. NA49 and NA61/SHINE have published several relevant data~\cite{na49Antiprotons, Aduszkiewicz:2017sei} that are used for tuning cosmic-ray formation and propagation models. The measurement of light nuclei in various $A$--$A$ data sets can be used to study the production of light ions at the threshold. These measurements will complement the NA49~\cite{dbarna49, Anticic:2016ckv} and ALICE results and allow to test the coalescence and thermal models in a different regime. Also, a first pilot run of carbon fragmentation measurement was conducted by the end of 2018 and demonstrated that the measurements are possible~\cite{2019arXiv190907136U}. Extended data taking with an upgraded NA61/SHINE experiment is planned after the CERN Long Shutdown 2.

\paragraph{Apparatus for Meson and Baryon Experimental Research (AMBER)}

AMBER will be the next-generation successor of the CERN SPS COMPASS experiment. It will perform measurements with protons between 50 and 280\,GeV/$c$ on fixed liquid hydrogen and helium targets~\cite{Denisov:2018unj}. The experiment is a magnetic spectrometer consisting of a number of subdetectors for the particle identification (including ring image Cherenkov, electromagnetic and hadronic calorimeters, gas electron multipliers). 
For 20\,bins in momentum from 10 to 50\,GeV/$c$ and 20\,bins in transverse momentum with 75\% beam purity at $5\times10^5$ protons-per-second beam intensity, it is expected that the statistical error is on the percent level for most of the differential cross section bins within several hours of beam time for each energy.

\paragraph{A Large Ion Collider Experiment (ALICE)}

ALICE is an LHC experiment that uses specific energy loss, time-of-flight, transition radiation, Cherenkov radiation, and calorimetric measurements for the particle identification~\cite{Abelev:2014ffa}. ALICE data are already actively used for constraining antinuclei production models, and light antinuclei production studies will continue in the following years. Furthermore, ALICE recently published interaction cross section measurements of the produced antinuclei with the detector materials~\cite{ALICE:2020zhb}. Very little data existed in this regard before, and they were already used to update the predictions for cosmic antideuterons~\cite{Serksnyte:2022onw}.
Motivated by~\cite{Winkler:2020ltd, kachelriess2021comment, winkler2021response}, ALICE also started to explore if it is possible to measure the production cross section of antihelium-3 in $\bar\Lambda_b$ decays. Including this production channel for antihelium-3 might boost the dark-matter-induced antihelium flux.

\paragraph{Large Hadron Collider Beauty (LHCb)}

In addition to the antiproton production cross section measurements mentioned above, LHCb aims at antideuteron production cross section measurements in $p$--$p$ interactions. LHCb is unique because it measures in the very forward direction ($2<\eta<5$). Particle identification uses a vertex locator around the collision point, ring imaging Cherenkov detectors, electronic and hadronic calorimeters, tracking stations, and muon stations. Antideuterons can be measured at LHCb in prompt production in $p$--$p$ collisions, in decays of heavy-hadrons, and fixed-target collisions. Also, LHCb started analyzing the production cross section of antihelium-3 in $\bar\Lambda_b$ decays.

\pagebreak

\bibliographystyle{naturemag_noURL}

\bibliography{main.bib}

\end{document}